\documentclass[aps,pre,twocolumn,groupedaddress,floatfix,showpacs,keywords]{revtex4}

\usepackage{graphicx}
\usepackage{amssymb}

\begin{document}

\title{Scaling and dynamics of sphere and disk impact into granular media}

\author{Daniel I. Goldman}
\email[]{daniel.goldman@physics.gatech.edu}
\homepage{http://www.physics.gatech.edu/goldman/}
\affiliation{School of Physics,
                Georgia Institute of Technology,
                Atlanta, GA 30332}
\author{Paul Umbanhowar}
\affiliation{Department of Mechanical Engineering,
Northwestern University, Evanston, IL 60208}

\date{\today}

\begin{abstract} Direct measurements of the acceleration of spheres and disks impacting granular media reveal simple power law scalings along with complex dynamics which bear the signatures of both fluid and solid behavior. The penetration depth scales linearly with impact velocity while the collision duration is constant for sufficiently large impact velocity. Both quantities exhibit power law dependence on sphere diameter and density, and gravitational acceleration.  The acceleration during impact is characterized by two jumps: a rapid, velocity dependent increase upon initial contact and a similarly sharp, depth dependent decrease as the impacting object comes to rest.  Examining the measured forces on the sphere in the vicinity of these features leads to a new experimentally based granular force model for collision.  We discuss our findings in the context of recently proposed phenomenological models that capture qualitative dynamical features of impact but fail both quantitatively and in their inability to capture significant acceleration fluctuations that occur during penetration and which depend on the impacted material.
\end{abstract}


\maketitle

\section{Introduction}  Collisions with complex particulate materials occur in diverse situations ranging from asteroid impact~\cite{meloshbook} to the penetration of a running crab's leg into beach sand~\cite{burAhoy73}.  Accordingly, collisions with granular media have long been investigated ~\cite{robins}, and like many areas of granular research, are being actively explored today in experiment~\cite{picAlar,debAwal04,allAmay57a,forAluk92,lohAber04,houApen05,bogAdra96b}, simulation~\cite{picAlar,wadAsen06}, and theory~\cite{allAmay57a,allAmay57b,tsiAvol05,ambAkam05,forAluk92,bogAdra96b,debAwal04}.  However, because the physics of such events must account for both fluid- and solid-like behavior during impact, understanding remains limited. No comprehensive continuum theory exists for even the relatively low impact velocity of a rock dropped into beach sand from an outstretched hand.

\begin{figure}[h!tb]
\begin{center}
\includegraphics[width=3in]{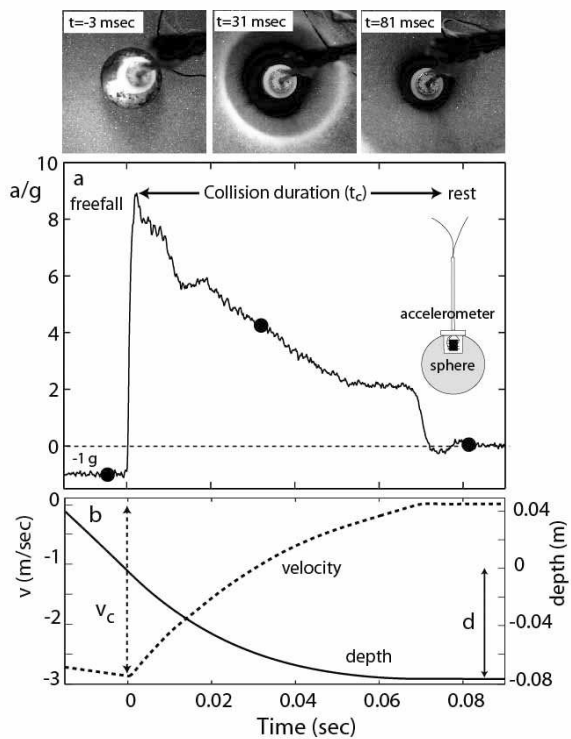}
\caption{Three sequential images (overhead view) of an $R=1.91$~cm steel sphere impacting glass beads at $v_c=2.86$~m/sec with (a) the acceleration and (b) the velocity and penetration depth of the sphere during the collision (the three dots ($\bullet$) in (a) correspond with the images). The depth is defined as the distance from the lowest point on the sphere to the initial free surface of the grains. Also in (a) is a sketch of the instrumented projectile (not to scale) showing a single-axis accelerometer embedded in a sphere.}
\label{figure1}
\end{center}
\end{figure}

Recent experiments and simulations of low velocity (below 5~m/sec) impact with granular media have mainly considered the depth to which an object penetrates before stopping. These studies have investigated how the penetration depth scales with various system parameters.  Durian's group \cite{ambAsan05,ambAkam05} performed experiments at low collision velocities $v$ (maximum penetration depth was approximately a sphere diameter) on a variety of spheres of differing radii $R$, sphere density $\rho_s$, and granular particle density $\rho_g$ and found that the penetration depth $d$ scaled as $d \sim v_c^{2/3} (\frac{\rho_s}{\rho_g})^{1/2} R^{2/3}$. de~Bruyn's group \cite{debAwal04} found a different scaling: $d \sim v_c^{1} (\frac{\rho_s}{\rho_g})^{1/2} R^{1/2}$. The latter experiments were conducted at larger impact velocities and with spheres of higher density such that maximum penetration depths were much greater than the sphere diameter $d \gg 2R$.  In two-dimensional (2D) disk simulations, Tsimring's group found that $d \sim v_c^{4/5} (\frac{\rho_s}{\rho_g})^{2/5} R^{3/5}$.

Phenomenological models have been proposed to account for experimental results of penetration depth from spheres impacting granular materials in a gravitational field~\cite{tsiAvol05,ambAkam05,forAluk92,bogAdra96b,debAwal04}. In all models, the force on a vertically falling object impacting a horizontal granular medium is written in the general form
\begin{equation}
m \frac{d^2 z}{dt^2}=-mg + F_d,
\end{equation}
where $m$ is the mass of the impactor, $z$ is the displacement of the lowest point on the object below the initial free surface of the grains, $g$ is the gravitational acceleration, and $F_d$ is the drag force due to the presence of the granular medium. Impact models typically represent the drag force as the sum of two terms:
\begin{equation}
\label{tmodel}
F_d=F_z+\alpha v^2,
\end{equation}
where $v=dz/dt$, and $F_z$ is posited to be a frictional/hydrostatic term with different proposed forms~\cite{tsiAvol05,ambAkam05} depending on the impact regime, and impactor geometry. For example \cite{tsiAvol05}, assumes that an impacting sphere experiences hydrostatic-like forces, basing their arguments on experiments of ~\cite{stoAbar04} in which the force on a flat intruder moving at a low constant velocity increases linearly with depth below the free surface. They argue that $F_z$ for shallow impact increases as $F_z=\eta \rho_g g z^2 D$ with $\eta$ a parameter dependent on grain properties (e.g.\ angle of repose), and $\rho_g$. For deep impact, the free surface is assumed to move along with the sphere such that the bottom of the object is always a fixed distance $z_0$ below the local free surface; thus $F_z$ approaches a constant with value $F_z=\eta \rho_g g z_{0}^2 D$.

Durian's group~\cite{ambAkam05} fitted penetration depth scaling data from experiment to derive a functional form for the frictional drag in the shallow penetration regime such that,
\begin{equation}
\label{dmodel}
F_z=mg + mg[3(z/d_0)^2-1] \exp (-2 |z|/d_1),
\end{equation}
where $d_0$ is the penetration depth for an object impacting the surface with initial collision velocity $v_c=0$, and $d_1=m/ \alpha$. These two parameters are independent of $v_c$. Similar to the model of Tsimring and Volfson~\cite{tsiAvol05}, Eq.\ [\ref{dmodel}] also scales like $z^2$ for $z \ll D$, and approaches a constant ($mg$) for sufficiently large impact depth. Recently Durian's group \cite{katAdur07} has proposed that a linear $F(z)$ (similar to the model proposed in \cite{lohArau}) provides a better fit.

To summarize, the granular medium is modeled by a force law with a hydrodynamic drag term proportional to the square of the velocity (which dominates at high velocity and thus deep penetration) and a term which accounts for a depth dependent static resistance force which dominates at low speeds and thus shallow penetration depths. Other studies have proposed that the drag force also includes a term linear in velocity~\cite{allAmay57a,allAmay57b,debAwal04}; however, the experiments in \cite{allAmay57a,allAmay57b} are in a much higher velocity regime, 700~m/sec, than the regime examined by us and in the other studies cited here where impact velocities are typically less than 5~m/sec.

Since at least 1742~\cite{robins} various force laws for granular impact have been proposed and their associated penetration scalings discussed; however, there have been no detailed experimental three-dimensional (3D) studies of the {\em forces} that the colliding object experiences during impact.  Measuring and understanding the forces exerted during impact is clearly important as demonstrated by the recent and surprising experimental finding that a disk comes to rest in a time $t_c$ independent of the initial impact velocity~\cite{picAlar}.  Fits of position versus time indicated that the acceleration during penetration was constant with magnitude dependent only upon $v_c$.

Accordingly, we describe here {\em direct} measurements of the forces exerted on a sphere during penetration of a granular medium. Integrated force measurements show how the penetration depth scales in deep impact experiments (our data most closely follow de~Bruyn's scaling~\cite{debAwal04}) while a systematic study of collision duration reveals simple scaling with system parameters.  We examine key features of the dynamics using the acceleration data and then use them to guide an examination of the forces at both high and low depths and velocities.  Constrained by experimental force data we propose an equation to describe the forces during granular impact.  Finally, we show that while our new impact force equation and those referenced above account for some of the features we observe, a wealth of dynamics associated with force fluctuations remains to be understood.

\section{Experimental details}

Figure~\ref{figure1} shows a sketch of the experimental apparatus. An MEMS IC accelerometer (Analog Devices ADXL150) with a range of $\pm 50$ $g$'s was mounted on an aluminum plug which was inserted into a hole or glued to a flat on top of the impactor; a small diameter tube isolating the accelerometer wires from the granular medium. We note that small tube does not affect the impact dynamics because the collapsing crater does not contact the tube until well after the impactor has stopped (see top panels of Fig.~\ref{figure1}~\cite{thoAshe}.) We choose acceleration to be positive in the direction opposite gravity. Impactors were dropped into the granular medium from heights of 0.01~m to $\sim 2.5$~m with corresponding impact velocities $v_c$ ranging from 0.4 to 7~m/s.

A variety of impactors and granular materials were used, see Tables \ref{materials} and \ref{impactors}.  Since holes were drilled in some spheres while flats were made on others and the mass of the accelerometer is included in the indicated sphere mass, the ``effective" sphere density, $\rho_s$, calculated using the masses and radii in the tables does not necessarily match the density of the material of which the sphere is composed.  Additionally, six 1.3~cm radius brass cylinders with different masses were mounted atop the 1.9~cm radius nylon sphere, which varied $\rho_s$ from 1.88 to 9.31~grams/cm$^3$.  Various containers were also employed (see Table~\ref{containers}), including a 29~cm diameter by 40~cm high PVC bucket, a 25~cm diameter by 30~cm high aluminum pot, a 50~cm diameter by 75~cm high cardboard barrel, and a 10~cm diameter by 10~cm high glass jar. Two different procedures were used to prepare the granular material:  1)  The container was vigorously rocked from side to side with decreasing amplitude until the surface was level.  2)  A sieve with outer diameter approximately equal to the container was placed in the bottom of the container, the material poured in, and the sieve pulled slowly to the surface. Both procedures produced reproducible dynamics.  The majority of experiments were conducted with bronze, steel, or nylon spheres or the bronze disk impacting the 0.25-0.42~mm glass beads prepared via rocking in the 30~cm diameter PVC container filled to a depth of approximately 25~cm.  Additionally, to vary the effective gravitational acceleration during impact, an Atwood machine was used to drop the bucket containing the granular material with accelerations ranging from zero to nearly $-g$.

\bigskip
\begin{table}[ht!]
\caption{Granular Media
\label{materials}}
\begin{tabular}{lccr}
  \hline
  \hline
  Material & Size (mm) & $\rho_{\mathrm{bulk}} (\mathrm{grams/cm}^3)$ & $\theta_r$ \\
  \hline
  Glass spheres& 0.25-0.42 & 1.56 & $23^\circ$\\
  Aluminum shot& $1\times1$ & 1.62 & $31^\circ$\\ 
               &(dia.$\times$len.)& &\\
  Millet seed & 1.2 & 0.72 & $28^\circ$\\ 
  Bronze spheres& 0.05 \& 0.17 & 5.49 & $24^\circ$\\ 
  \hline
  \hline
\end{tabular}
\end{table}

\begin{table}[ht!]
\caption{Impactors
\label{impactors}}
\begin{tabular}{lcr}
  \hline
  \hline
  Impactor & Radius (cm) & Mass (grams) \\
  \hline
  Steel Sphere& 9.5 &  34\\
\hspace{0.25 in}'' & 1.3  & 83\\
\hspace{0.25 in}'' & 1.5 &  130\\
\hspace{0.25 in}'' & 2.0 &  287\\
\hspace{0.25 in}'' & 2.5 &  531 \\
\hspace{0.25 in}'' & 3.5 &  1437\\
\hspace{0.25 in}'' & 4.0 &  2099\\
\hspace{0.25 in}'' & 4.5 &  3055\\
\hspace{0.25 in}'' & 5.0 &  4079\\
  Bronze Sphere & 1.3 & 64\\
\hspace{0.25 in}'' & 1.9 & 201\\
\hspace{0.25 in}'' & 2.6 &  518 \\
  Nylon Sphere & 1.25 &  9 \\
\hspace{0.25 in}'' & 1.9 &  18 \\
  Bronze Disk & 1.0 &  10 \\
  (5 mm high) &  &  \\
  \hline
  \hline
\end{tabular}
\end{table}

\begin{table}[ht!]
\caption{Containers
\label{containers}}
\begin{tabular}{l p{.6in} p{.5in} p{.7in} l}
  \hline
  \hline
 Container & Diameter (cm) & Height (cm) &{\raggedleft Wall Thickness (cm)}\\
\hline
Cardboard Barrel& 50 & 75 & 0.3\\
PVC Bucket& 28.5 & 38  &  0.3 &\\
Aluminum Pot& 25 & 30 & 0.33\\
\hline
\hline
\end{tabular}
\end{table}

\section{Impact and derived quantities}

Figure~\ref{figure1} shows the acceleration, velocity, and position of a sphere colliding with a granular medium. We measure the acceleration directly and integrate to obtain position and velocity. At impact, the acceleration increases rapidly as the material suddenly applies force. The acceleration decreases as the sphere penetrates into the medium and finally comes to rest at a finite penetration depth $d$ in a time $t_c$. We discuss the details of the acceleration profile in Section IV. We begin with a discussion of the dependence of $d$ and $t_c$ on the initial impact velocity $v_c$ as well as sphere density $\rho_s$, sphere radius $R$, and gravitational acceleration $g$.

\subsection{Penetration depth}
\label{penDepth}

\begin{figure}[h!tb]
\begin{center}
\includegraphics[width=3in]{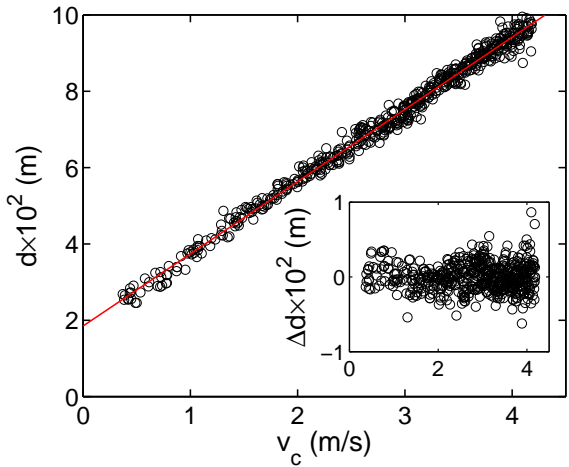}
\caption{Penetration depth vs.\ impact velocity for a bronze sphere ($R=1.91$~cm) impacting glass beads. Solid line is a fit with $d=(\Delta d/\Delta v_c) v_c + d_0$ with $\Delta d/\Delta v_c = 0.0189$~sec and $d_0 = 0.0185$~m. Inset: residuals of the fit.}
\label{depthvsvel}
\end{center}
\end{figure}

Previous studies~\cite{debAwal04,ambAkam05,tsiAvol05} have discussed the dependence of the penetration depth $d$ on the impact velocity or, equivalently, the total change in potential energy of the impactor (refer to the discussion in the introduction for scalings obtained in this earlier work). Figure~\ref{depthvsvel} shows that for large enough $v_c$, such that a sphere penetrates more than approximately its radius, $d$ increases linearly with $v_c$. The linear scaling agrees with the data and the scaling proposed in~\cite{debAwal04}. We do not investigate shallow impact~\cite{ambAkam05} in which this scaling is expected to be modified~\cite{tsiAvol05} due to the varying effective cross section of the sphere for $z < R.$

\begin{figure}[h!tb]
\begin{center}
\includegraphics[width=3.25in]{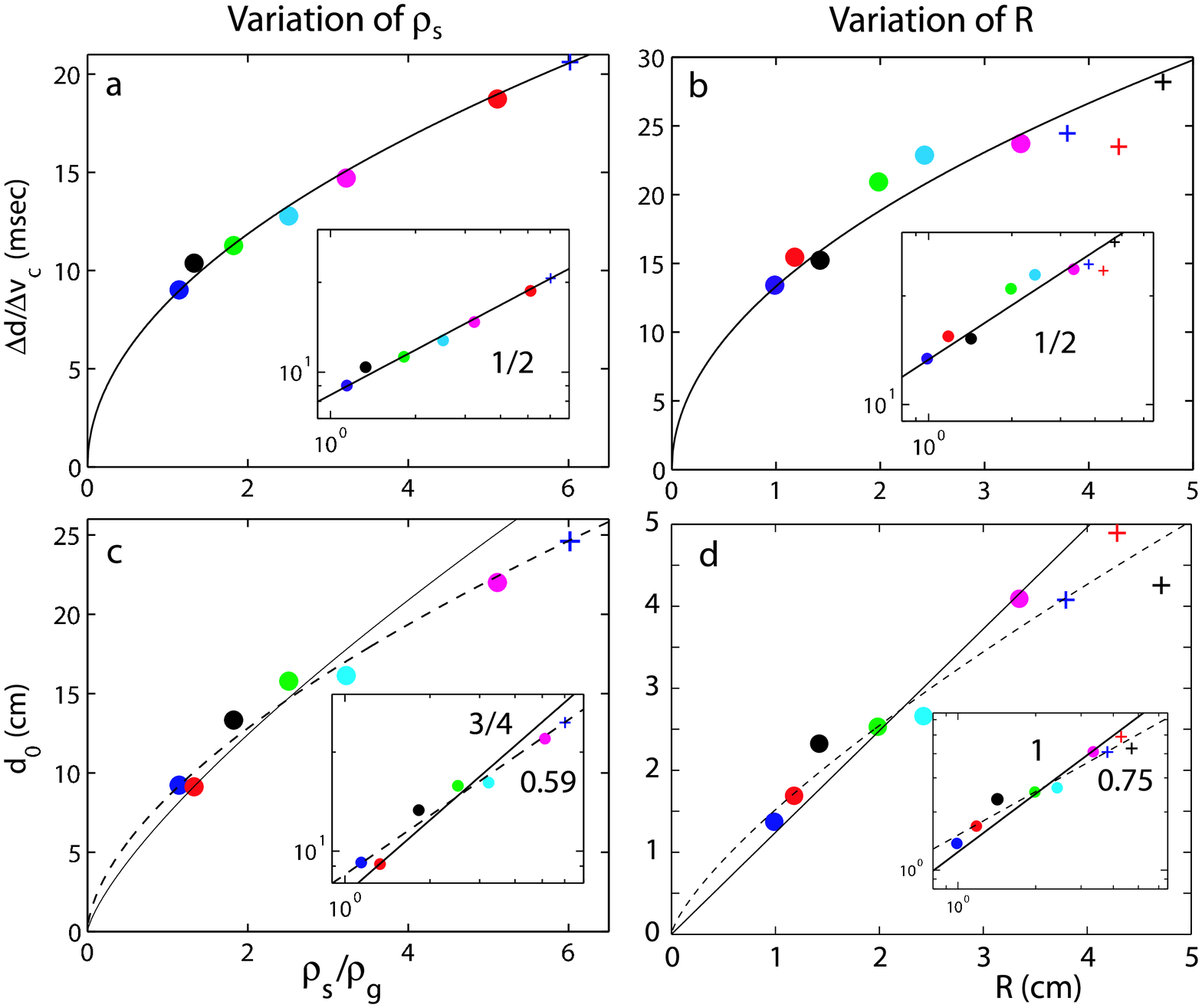}
\caption{(a, b) Slope and (c, d) intercept of linear fits of penetration depth vs.\ sphere impact velocity in glass beads for a 1.9~cm nylon sphere (left column) with $\rho_s = 1.88,$ 2.08, 2.85, 3.91, 5.03, 7.97, and 9.38~grams/cm$^3$ (blue, red, black, green, cyan, magenta $\circ$ and blue $+$) and steel spheres (right column) with  $R=0.95,$ 1.27, 1.51, 1.98, 2.46, 3.49, 3.97, 4.52, and $5.00$~cm and corresponding masses $m=34.23,$ 66.3, 112, 287, 531, 1437, 2099, 3055 and 4079~grams (blue, red, black, green, cyan, magenta $\circ$ and blue, red, black $+$). The solid curves in (a, b) are proposed scalings from~\cite{debAwal04} with $d - d_0 \sim (\rho_s/ \rho_g)^{1/2}$ and (c, d) from \cite{ambAkam05}  with $d_0 \sim (\rho_s/\rho_g)^{3/4} R.$  In (c, d), the dashed curves are power law fits with exponents $0.59$ and $0.75$ respectively.}
\label{depthvsvarious}
\end{center}
\end{figure}

We systematically vary $R$ and $\rho_s$ to determine how penetration depth changes with these parameters. Since our data indicates the dependence of $d$ on $v_c$ is  close to linear for $d>R$, such that $d=d_0+ v_c \Delta d/\Delta v_c$, we compute the slope, $\Delta d/\Delta v_c$, and intercept, $d_0$, of $d$ vs.\ $v_c$ and plot these as functions of $\rho_s$ and $R$ as shown in Fig.~\ref{depthvsvarious}. We compare our results for $\Delta d/\Delta v_c$ vs.\ $v_c$  to the scaling proposed by de~Bruyn~\cite{debAwal04}, $d \sim (\rho_s/\rho_g)^{1/2}R^{1/2} v_c$, and our results for $d_0$ vs.\ $v_c$ to the findings of Ambroso~{\em et al.}~\cite{ambAkam05}, $d_0 \sim (\rho_s/\rho_g)^{3/4} R,$ since the former reference makes no predictions for the scaling of $d_0$. As Fig.~\ref{depthvsvarious}(a, b) demonstrates, the slope of $d$ vs.\ $v_c$ scales with the sphere density as $d \sim (\rho_s/ \rho_g)^{1/2}$ and with the sphere radius as $d \sim R^{1/2}$ as in~\cite{debAwal04}.   Figure \ref{depthvsvarious}(c, d) indicates that better fits (dashed curves) for the scaling of $d_0$ as a function of $\rho_s/\rho_g$ and $R$ are obtained with exponents of 0.59 and 0.75 respectively as opposed to the predicted values (solid curves) of 0.75 and 1.  However, comparison of the measured and predicted values of the exponents is not strictly valid since our fits apply to the $d>R$ region where $d \propto v_c$ whereas those in \cite{ambAkam05} are for $d<R$ and give the actual penetration depth of a sphere released at the surface ({\em i.e.}\ $v_c=0$).

\begin{figure}[h!tb]
\begin{center}
\includegraphics[width=3.5in]{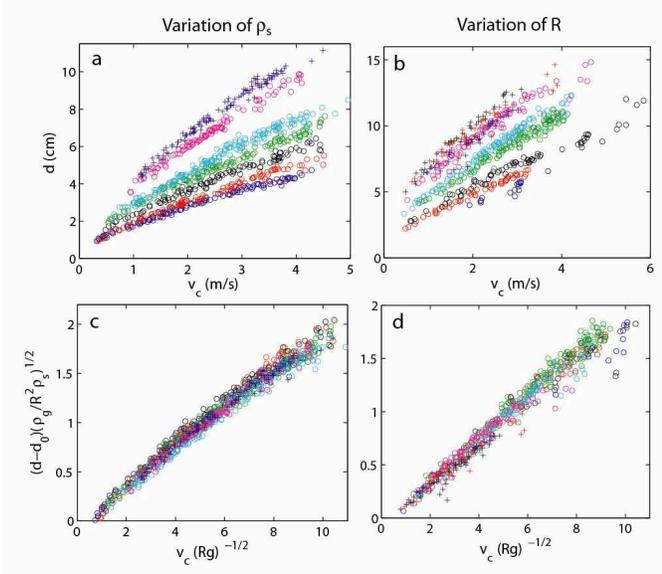}
\caption{Raw data and collapse using the scalings from Fig.~\ref{depthvsvarious}. First column (a,c): Penetration depth and scaled penetration depth vs.\ $v_c$ for varying $\rho_s$. Second column (b,d): Penetration depth and scaled penetration depth vs.\ $v_c$ for varying $R$.  Symbols are the same as in Fig.~\ref{depthvsvarious}. }
\label{depthscaling4panel}
\end{center}
\end{figure}

Combining these scalings, we write the penetration depth for $d > R$ as
\begin{equation}
\label{dscale}
d = C_1 v_c \sqrt{\frac{R \rho_s}{g \rho_g}} + C_2 (\rho_s/\rho_g)^{0.59} R^{0.75},
 \end{equation}
 where $C_1$ and $C_2$ are constants.  Figure~\ref{depthscaling4panel} shows both the unscaled penetration depth data and the collapse obtained using Eq.~\ref{dscale}. Although we did not systematically investigate the influence of particle (grain) diameter, $r,$ on penetration depth, Eq.~\ref{dscale} suggests  $\Delta d/\Delta v_c \sim r^{0}$ and $d_0 \sim r^{1/4}$.

\begin{figure}[h!tb]
\begin{center}
\includegraphics[width=2.5in]{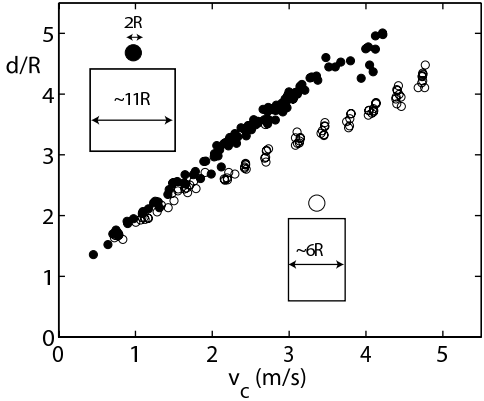}
\caption{Container diameter affects penetration depth. A steel sphere with $R=2.5$~cm  impacting glass beads in cylindrical containers with diameters $28$~cm ($\bullet$) and $14.6$~cm ($\circ$). Both containers are filled to a depth of approximately 30~cm. The data ($\bullet$) is reproduced from Fig.~\ref{depthscaling4panel}. The sketches show the relative sphere and container sizes.}
\label{depthvscontainer}
\end{center}
\end{figure}

Figure~\ref{depthvscontainer} shows that the penetration depth is affected by finite container size.  Decreasing the container diameter by approximately a factor of two decreases the slope of $d$ vs.\ $v_c$ by about 1.3 for $v_c \gtrapprox 2$~m/s. There is little apparent difference in $d$ for smaller $v_c.$  Accordingly, the apparent sublinear dependence of $d$ on $v_c$ evident in the collapsed data in Fig.~\ref{depthscaling4panel}(c) might result from the finite diameter of the container, which, as Fig.~\ref{depthvscontainer} indicates, becomes more significant the further the impactor penetrates.

\subsection{Collision duration, $t_c$}

\subsubsection{Independence of $t_c$ on $v_c$}

\begin{figure}[h!tb]
\begin{center}
\includegraphics[width=3.25in]{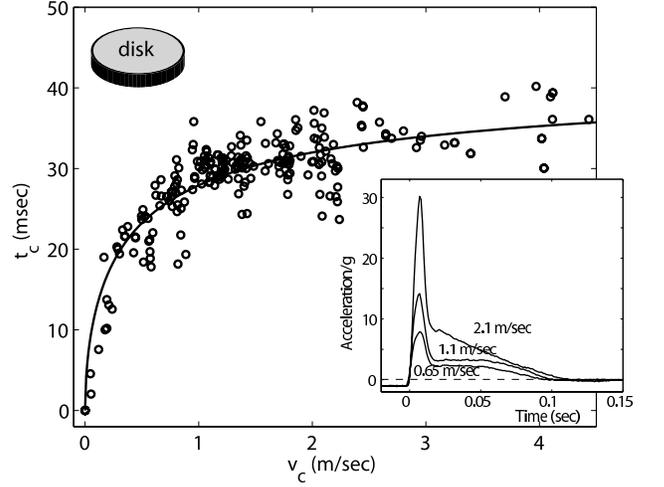}
\caption{Collision duration for impact of an $R=1.0$~cm, $m=0.01$~kg disk with glass beads. Solid line shows fit from the model~\cite{tsiAvol05} proposed for large impact velocity (deep penetration) with $\alpha=0.032$~kg/m and $F_z=0.112$~N.  At low velocity $t_c$ increases with $v_c$ which is opposite of spheres (see Figs.~\ref{tc_v} and \ref{tc_all}).}
\label{tc_disk}
\end{center}
\end{figure}

In a previous quasi-2D study of disks impacting smaller disks confined between narrow sidewalls Ciamarra {\em et al.\ } found that the collision duration $t_c$ was independent of $v_c$ ~\cite{picAlar}. They attributed this to a constant acceleration during impact whose magnitude was linearly dependent on $v_c$. Our acceleration data for a 3D disk reveals that $t_c$ (defined here as the time from impact to arrest, see {\em e.g.}\ Fig.~\ref{figure1}) is nearly independent of $v_c$ for $v_c \gtrsim 1.5$~m/s, see Fig.~\ref{tc_disk}. We denote this velocity independent collision duration as $t_0$. Unlike in~\cite{picAlar}, but as is the case for spheres, we find that the acceleration is not constant during the penetration phase (see inset of Fig~\ref{tc_disk}); we return to analysis of the acceleration in Section IV. Below $v_c \approx 1.5$~m/sec, $t_c$ decreases and as $v_c \rightarrow 0$, $t_c \rightarrow 0$. This decrease at low velocity is a consequence of the finite yield stress of granular materials (incorporated in the model of de~Bruyn~\cite{debAwal04}), which allows support of a distributed finite load without penetration~\cite{nedderman}.

We now make an argument for the dependence of $t_c$ on $v_c$ for disk impact based on the model of~\cite{tsiAvol05} in the deep collision regime. Using Eqs. (1) and (2) with constant $F_z$, appropriate for the regime of deep impact velocity ~\cite{tsiAvol05}, we write
\begin{equation}
dt=\frac{dv}{(F_z/m-g)+\frac{\alpha}{m} v^2}.
\end{equation}
We supplement this equation with the boundary condition for a disk such that $v(t=0)=-v_c$. Integration yields $t=-\frac{\tan^{-1}{\sqrt{\alpha/(F_z-mg)}v}}{\alpha (F_z/m^2-g/m)}+C_1$ with $C_1=\frac{\tan^{-1}{\sqrt{\alpha/(F_z-mg)}v_c}}{\alpha (F_z/m^2-g/m)}$.

\bigskip
We define the collision time $t_c$ as when the object comes to rest, $v=0$. Solving for this time yields

\begin{equation}
t_c=\frac{\tan^{-1}{\sqrt{\alpha/(F_z-mg)}v_c}}{\alpha (F_z/m^2-g/m)},
\end{equation}

This solution is plotted in Fig.~\ref{tc_disk} and is a good fit to the experimental data; $\alpha$ and $F_z$ are fit parameters. As $v_{c} \rightarrow \infty$, $t_c$ approaches $t_0=\pi/2 \sqrt{\frac{m}{\alpha(F_z/m-g)}}$, a constant independent of $v_c$.

The asymptotic form of $t_c$ given above allows a prediction of scaling behavior of $t_0$ with experimental parameters. Using $F_z \sim \rho_g g R^3$ and $\alpha \sim \rho_g R^2$ as predicted in ~\cite{tsiAvol05,allAmay57a}, $t_0$ should scale as

\begin{equation}
t_0 \sim \mathrm{C} (\frac{\rho_s}{\rho_g}) \sqrt{R/g}
\label{proposedt0}
\end{equation}
 where $\mathrm{C}$ is a constant that could depend on impactor and grain geometry, density, friction coefficients, normal dissipation and other material parameters.

\subsubsection{Scaling of $t_0$ in sphere data}

\begin{figure}[h!tb]
\begin{center}
\includegraphics[width=3in]{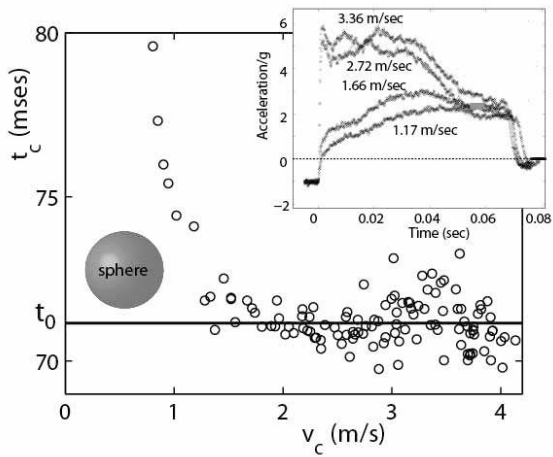}
\caption{Collision time $t_c$ is independent of collision velocity $v_c$ for sufficiently high $v_c$ ($R=2.0$~cm steel sphere into glass beads). Inset: acceleration vs.\ time for four impact events showing that while $t_c$ is independent of $v_c$, the acceleration profile is not.}
\label{tc_v}
\end{center}
\end{figure}

The experimental result for the disk data showing that $t_c \rightarrow t_0$ as $v_c$ increases is also seen in our sphere data, see Fig.~\ref{tc_v}. Similarly, the acceleration during collision is not constant, see inset of Fig.~\ref{tc_v}. The large range in system parameters varied in the sphere data set allow us to examine for the first time how $t_0$ scales with radius, density and gravitational acceleration.

In contrast to the disk, in which $t_c \rightarrow 0$ as $v_c \rightarrow 0$, $t_c$ {\em increases} with decreasing impact velocity for spheres; compare Figs.~\ref{tc_disk} and \ref{tc_v} for $v_c  \lessapprox 1.5$~m/sec. We attribute this difference to the fact that even as  $v_c \rightarrow 0$, the small initial contact area between sphere and grains due to the curvature of the sphere always produces local stresses sufficient for grain bed yielding, which consequently allows the sphere to penetrate for a finite time. Thus, a sphere, unlike a disk, always penetrates a finite distance into the material even with $v_c=0$; this penetration regime has been examined by \cite{ambAsan05, debAwal04}.

\begin{figure}[h!tb]
\begin{center}
\includegraphics[width=2in]{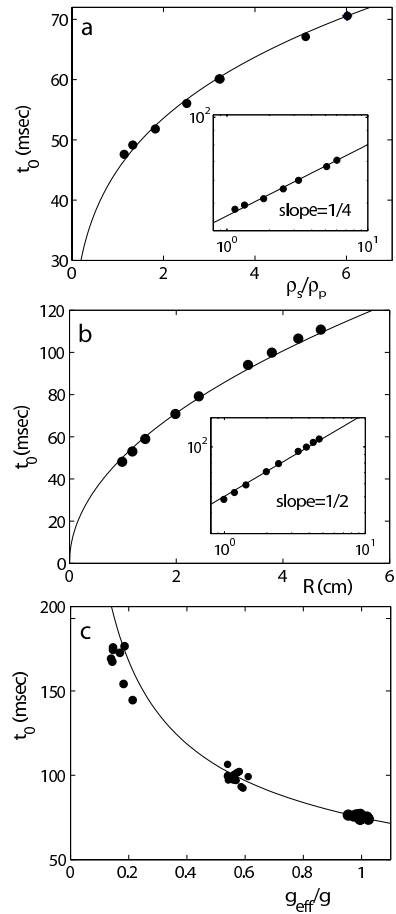}
\caption{Scaling of asymptotic ($v_c>1.5$~m/sec) penetration time $t_0$ for impact into glass beads as a function of (a) sphere density for the $R=1.9$~cm nylon sphere, fit shown with $t_0$ in seconds is $t_0=0.045 (\rho_s/\rho_g)^{1/4}$, (b) sphere radius for steel spheres, fit shown is $t_0 = 0.050 R^{1/2}$, and  (c) effective gravitational acceleration for an $R=1.98$~cm steel sphere, fit shown is $t_0 = 0.075 (g_{\tiny \mbox{eff}}/g)^{-1/2}$. }
\label{tc_scaling}
\end{center}
\end{figure}

For sufficiently high $v_c$, the surface of the sphere in contact with the granular medium is expected to be essentially constant for a large fraction of the collision interval after the
initial impact \cite{tsiAvol05}. We therefore expect the proposed scaling of Eq.~ \ref{proposedt0} for disks to be obeyed for spheres as well in the high $v_c$ regime. As shown in Fig.~\ref{tc_scaling}, we find that, $t_0 \sim R^{1/2}$, and $t_0 \sim (g_{\tiny \mbox{eff}})^{-1/2}$, where $g_{\tiny \mbox{eff}}$ is the acceleration of the falling bucket in the Atwood machine. For impact at varying sphere density, the figures shows that $t_0 \sim (\rho_s/\rho_g)^{1/4}$ which implies that $C$ in Eq.~\ref{proposedt0} should vary as $(\rho_s \rho_g)^{-1/4}$.

\begin{figure}[h!tb]
\begin{center}
\includegraphics[width=3.5in]{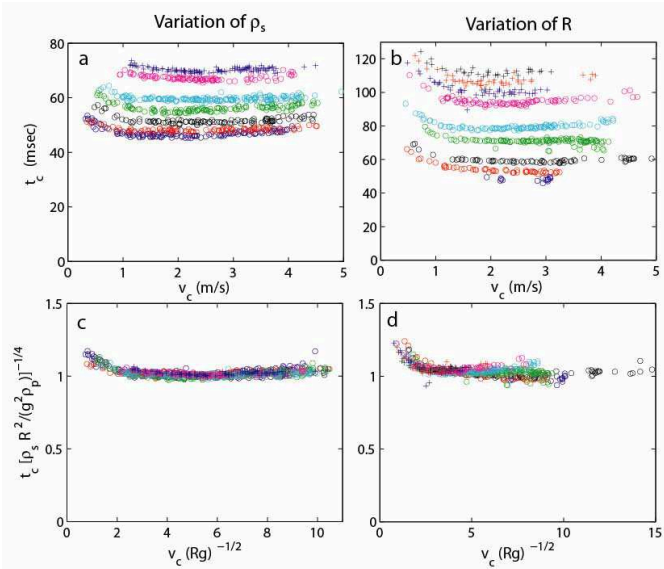}
\caption{Test of scaling derived from data in Fig.~\ref{tc_scaling}. Collision time vs.\ $v_c$ for (a, c) varying sphere density (nylon sphere with added masses) and (b, d) varying sphere radius (steel spheres).  The data in (c, d) are re-scaled data using Eq. (\ref{t0scaling}). The effective densities of the $R=1.91$~cm nylon sphere are $\rho_s=1.88$, 2.08, 2.85, 3.91, 5.03, 7.97, and 9.38~grams/cm$^3$, corresponding to blue, red, black, green, cyan, magenta $\circ$ and blue $+$. The radii of the steel spheres are $R=0.95$, 1.27, 1.51, 1.98, 2.46, 3.49, 3.97, 4.52, and 5.00~cm with associated masses ranging from $m=34$ to 4079~grams (blue, red, black, green, cyan, magenta $\circ$ and blue, red, black $+$).}
\label{tc_all}
\end{center}
\end{figure}

Combining these three scalings, we obtain
\begin{equation}
t_0 \sim (\frac{\rho_s}{\rho_g})^{1/4} \sqrt{R/g}
\label{t0scaling}
\end{equation}
or in terms of the sphere mass $t_0 \sim [M/(R g^2 \rho_g)]^{1/4}.$  A test of this final scaling is shown in Fig.~\ref{tc_all} which shows that the $t_c$ data collapses well for a large range of densities and radii.  Comparing the experimentally determined $t_0$ scaling to our model shows that Eqs.~\ref{t0scaling} and \ref{proposedt0} both include a $\sqrt{R/g}$ term but have different exponents in the density term ($1/4$ and $1$ respectively).

\section{Dynamical features of impact}

\begin{figure}[h!tb]
\begin{center}
\includegraphics[width=3.5in]{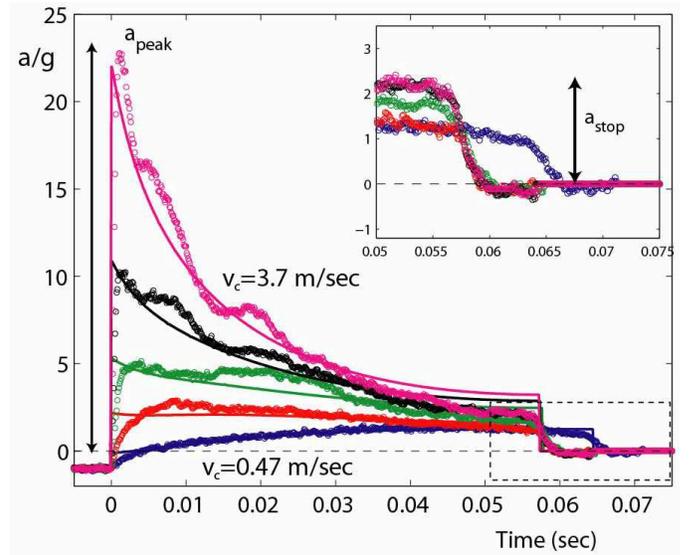}
\caption{(Color online) Acceleration profiles for different impact velocities ($v_c=0.47,0.93,1.99,2.88,3.72$~m/sec, blue, red, green, black, pink) for a 147~gram nylon sphere (R=1.9~cm) impacting polydisperse 0.25-0.42 mm glass beads. Dynamical features shown and discussed in the text are the peak acceleration $a_{\mathrm{peak}}$, the jump in acceleration as the sphere comes to rest, $a_{\mathrm{stop}}$ (see inset), and the time for the object to come to rest, $t_c$. The open symbols are experimental data, the solid lines are best fits using the model of Ambroso {\em et al}~ \cite{ambAkam05}; parameters of the fit are given in the text and in Fig.~\ref{d0d1}. The model fits are not shown in the inset.}
\label{AccProfileFitComp}
\end{center}
\end{figure}

\begin{figure}[h!tb]
\begin{center}
\includegraphics[width=3in]{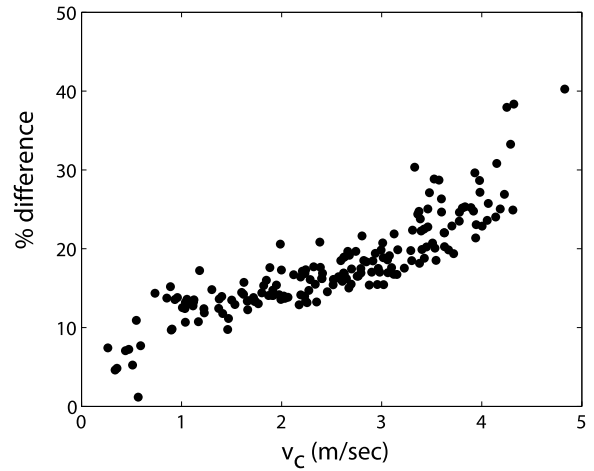}
\caption{The increasing deviation in average percent difference between experimental data and best fits to the model of~\cite{ambAkam05} (see Eq.~\ref{dmodel}) from the data set used in Fig.~\ref{AccProfileFitComp}. The \% difference calculation is averaged over the central 70 \% of the data and model fit.}
\label{modelfitdeviation}
\end{center}
\end{figure}

While $t_c$ and $d$ are characteristic physical quantities associated with any impact into deformable material, we hypothesize that the scaling that matches experimental data can be obtained with a variety of phenomenological models (see references in Section 1). In contrast, examine Fig.~\ref{AccProfileFitComp} which shows the acceleration profiles for impact of an $R=1.91$~cm diameter nylon sphere for five distinct impact velocities. As this figure makes clear, while current models capture the average physics involved in the impact events (we have fit the data to the model of ~\cite{ambAkam05}), they miss much of the detailed physics involved in a collision. As $v_c$ increases, this model fails to capture both significant acceleration fluctuations as well as the underlying form of the acceleration.  Figure~\ref{modelfitdeviation} shows the corresponding growth in the relative error between experimental data and the model fit as $v_c$ increases.

\begin{figure}[h!tb]
\begin{center}
\includegraphics[width=3in]{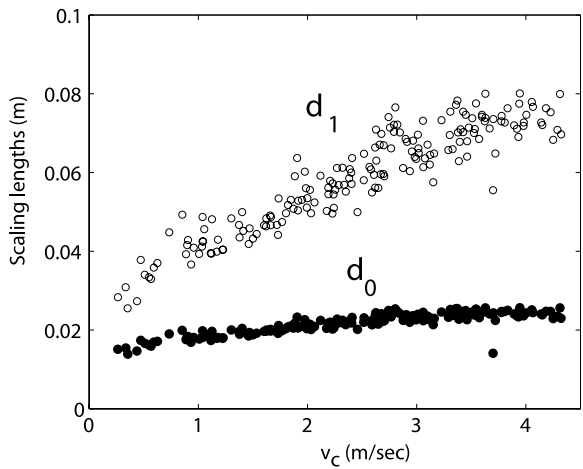}
\caption{\label{figure17} (a) The scaling lengths $d_0$ ($\bullet$) and $d_1$ ($\circ$) in the model of~\cite{ambAkam05} (see Eq.~\ref{dmodel}) as determined by best fits to the data set from Fig.~\ref{AccProfileFitComp} are not constants as predicted by the model but depend on the impact velocity.}
\label{d0d1}
\end{center}
\end{figure}

This model has a second shortcoming: it predicts that the scaling lengths $d_0$ and $d_1$ in Eq.~\ref{dmodel} are independent of $v_c$. Least squares fits of $a(t)$ for fixed sphere mass $M$ allowing $d_0$ and $d_1$ to vary are shown for different $v_c$ in Fig.~\ref{AccProfileFitComp}. The dependence of $d_0$ and $d_1$ on $v_c$ is shown in Fig.~\ref{d0d1}. Although $d_0$ is roughly independent of $v_c$, $d_1$ increases with increasing $v_c$. Since the model fails to account for the dependence of $d_0$ and $d_1$ on $v_c$, this indicates that additional physics is needed to fully characterize impact dynamics.

\begin{figure}[h!tb]
\begin{center}
\includegraphics[width=3in]{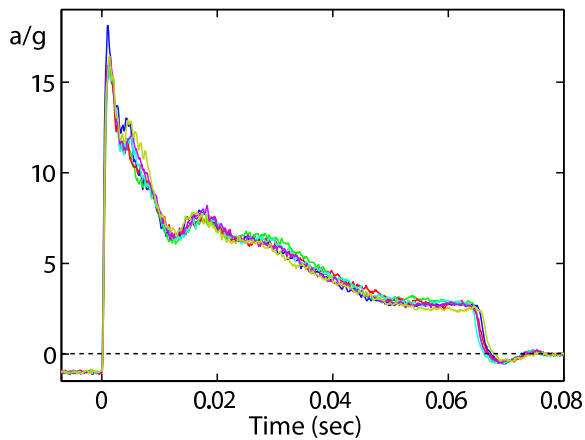}
\caption{(Color online) The details of the acceleration profiles show only small variation (net $<5\%$) from run to run for six separate impacts of a $R=1.91$~cm bronze sphere into glass with initial velocity $3.65 < v_c < 3.67$~m/s.}
\label{repeatable}
\end{center}
\end{figure}

Therefore, we now discuss the detailed acceleration profile of a sphere as it impacts a granular medium. We first describe two robust acceleration features seen for a wide range of impactor radii, densities, and material types, as well as for a range of granular materials: a peak in the acceleration during collision $a_{\tiny \mbox{peak}}$ and a rapid decrease in the acceleration as the object comes to rest, $a_{\tiny \mbox{stop}}$, see Fig.~\ref{AccProfileFitComp}. For impact at fixed parameters ({\em i.e.}\ $v_c$, $R$, $\rho_g$,$\rho_s$) Fig.~\ref{repeatable} shows that the acceleration profile and associated features are reproducible to within approximately $5\%$ from run to run. We demonstrate how these dynamical features scale with system parameters.  Using insights gained from the scaling, we deduce an empirical force model by examining the experimental data in the extremes of high velocity/shallow penetration and  low velocity/deep ($d>R$) penetration.

\subsection{Peak accelerations}

\begin{figure}[h!tb]
\begin{center}
\includegraphics[width=3in]{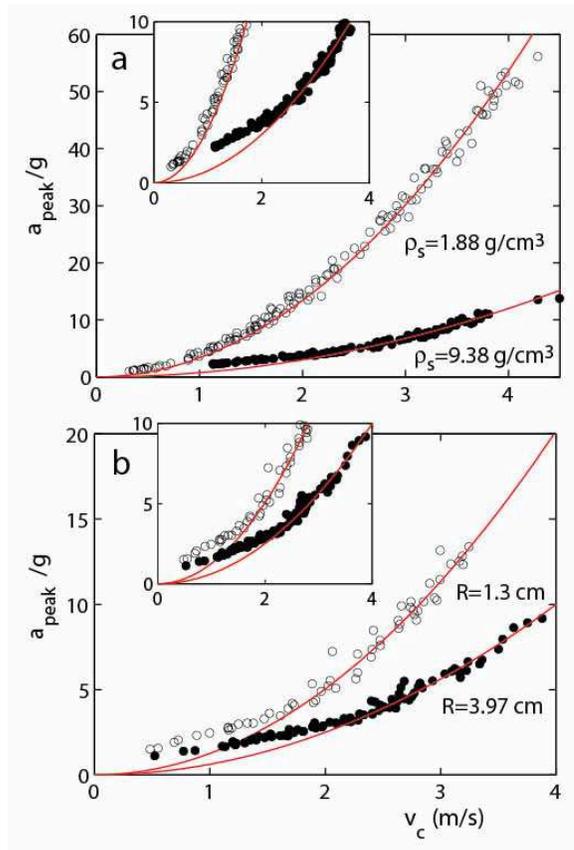}
\caption{(Color online) Peak impact acceleration $a_{\mathrm{peak}}$ of a sphere as a function of impact velocity into glass beads for (a) nylon sphere with effective density 1.88 and 9.38~grams/cm$^3$ and $R=1.9$~cm, and (b) steel spheres with radii $R=1.3$ and $3.97$~cm. Solid lines are fits of $a_{\mathrm{peak}} \sim v^2$ for $v_c \gtrapprox 1.5$~m/sec. The insets shows that as $v_c \rightarrow 0$, the $v_{i}^2$ scaling does not hold.}
\label{peak_accel}
\end{center}
\end{figure}

During the collision the acceleration rises to a maximum whose magnitude we denote $a_{\mathrm{peak}}$. At low velocity, the peak is not pronounced, but rather is seen as a broad maximum. As $v_{c}$ increases, the peak occurs soon after the initial contact of the sphere with the grain surface.  For $v_c \gtrapprox 1.5$~m/sec the interval between contact and peak acceleration is only a few milliseconds.  Figure~\ref{peak_accel} shows that in this regime, for all sphere densities and radii, $a_{\mathrm{peak}}$ increases approximately like $v_{c}^2$. $a_{\mathrm{peak}} \sim v^2$ is in accord with all models~\cite{ambAkam05,tsiAvol05,allAmay57a} in the high velocity and low depth limits, and implies that for these $v_c$ impact fluidizes grains sufficiently for the system to display inertial fluid-like drag~\cite{tritton89}.  For fixed $v_c$, $a_{\mathrm{peak}}$ can increase by more than 50\% if the container is sharply tapped a few times before impact.  But, as Fig.~\ref{repeatable} indicates, preparation by rocking produces consistent results.

At low impact velocity as $v_{c} \rightarrow 0$, $a_{\mathrm{peak}}$ does not approach zero, but instead approaches a finite intercept, see insets of Fig.~\ref{peak_accel}. Figure~\ref{AccProfileFitComp} shows that in the low velocity limit the peak acceleration occurs at the end of the collision just before the sphere comes to rest. We interpret this as a change from hydrodynamic-like dynamics at high velocities to a regime in which resistance to inertia is no longer the dominant source of drag.  This is consistent with Fig.~\ref{peak_accel}(a) which shows the lower density sphere scaling like $v_c^2$ over a larger velocity range than higher density sphere; the latter's velocity decreases more slowly resulting in deeper penetration where the influence of $F(z)$ is no longer negligible.  Since high and low velocity regimes are dominated by different physics, we postpone further discussion of $a_{\mathrm{peak}}$ until the treatment of force laws in Section~\ref{forceLaws}.

\subsection{Stopping acceleration}

For all $v_{c}$, the sphere does not come to rest gradually, but instead suffers an abrupt decrease in acceleration before halting, which is reminiscent of a horizontally sliding object stopping due to friction. We denote the magnitude of the decrease as $a_{\tiny \mbox{stop}}$. This impact feature occurs for all spheres, granular materials, and containers we employed in our study.  As can be seen in the inset of Fig.~\ref{figure1}, $a_{\tiny \mbox{stop}}$ increases with increasing $v_c$. The models of both Tsimring and Volfson~\cite{tsiAvol05}, and Ambroso {\em et al.}~\cite{ambAkam05} predict jumps in the acceleration as the impactor comes to rest (see Fig.~\ref{AccProfileFitComp}) which are attributed to the dominance of the depth dependent frictional/hydrostatic drag term, $F_z$. While the models predict $a(t)$ instantaneously jumps to zero at the end of penetration (Fig.~\ref{AccProfileFitComp}), we find instead that the rate at which $a(t)$ decreases to zero depends on $v_c$; as $v_c$ increases, the transition to $v=0$ sharpens (see inset of Fig.~\ref{AccProfileFitComp}). We comment on this in Section~\ref{forceLaws}.

\begin{figure}[h!tb]
\begin{center}
\includegraphics[width=3in]{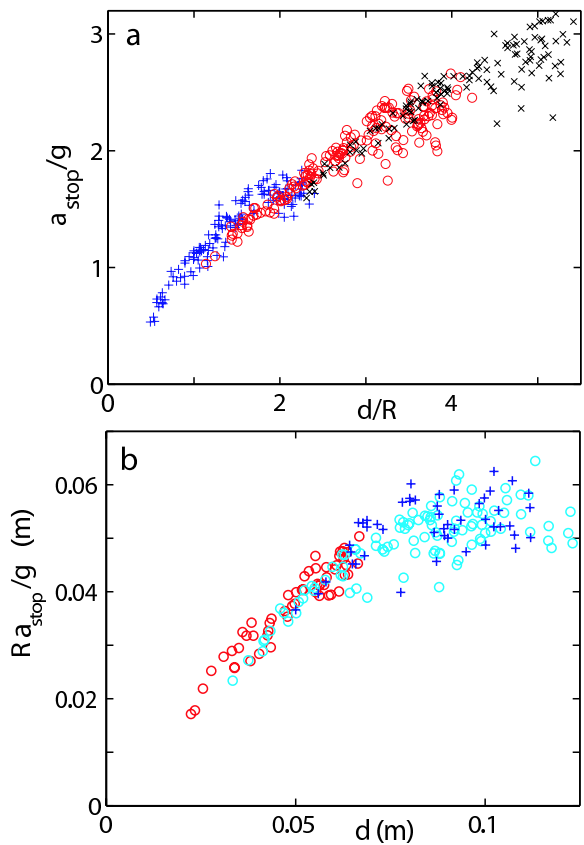}
\caption{(Color online) (a) Scaling of $a_{\tiny \mbox{stop}}$ vs.\ scaled impact depth for an $R=1.91$~cm nylon sphere with effective densities $\rho_s=1.78,$ 5.03, and 9.39~grams/cm$^3$ (blue +, red $\circ$, black $\times$) with corresponding masses $m=51.6,$ 146, and 272~grams.   (b) $a_{\mathrm{stop}}$ scaled by sphere radius vs.~impact depth for steel spheres of radii $R=1.3,$ 2.5, and 4.0~cm with masses $m=83.3$, 531 and 2099~grams (red, cyan, $\circ$ and blue, $+$).}
\label{astop_variation}
\end{center}
\end{figure}

In Fig.~\ref{astop_variation} we examine $a_{\tiny \mbox{stop}}$ as a function of penetration depth $d$ for varying sphere density $\rho_s$ and radius $R$, where $d$ is the depth to which the lowest point of the sphere penetrates the material (see Section~\ref{penDepth}). For varying $\rho_s$ [Fig.~\ref{astop_variation}(a)], we find that $a_{\tiny \mbox{stop}}$ is {\em independent} of the sphere density - at the same depth spheres of different density experience the same acceleration ({\em i.e.}\ force $\propto$ mass), further evidence for frictional forces dependent primarily on geometry dominating the final stages of penetration. $a_{\tiny \mbox{stop}}$ increases as the ultimate penetration depth of the sphere increases. For $d \gtrsim R$ $a_{\tiny \mbox{stop}}$ increases approximately linearly until $d/R \approx 4.5$ .  For $d \lesssim R$, $a_{\tiny \mbox{stop}}$ increases more rapidly with increasing $d$ (larger slope).   We attribute the increase in $a_{\tiny \mbox{stop}}$ with $d$ to an increase in the effective contact area between sphere and grains as the sphere stops with saturation indicating that the lower half of the sphere is fully in contact with solidified grains.  We return briefly to this point in Section \ref{forceLaws}.

For steel spheres with varying radii $a_{\tiny \mbox{stop}}$ also increases with depth and then saturates.   The saturation occurs at shallower scaled depth and smaller $a_{\tiny \mbox{stop}}$ 
for increasing $R$. Figure~\ref{astop_variation}(b) shows that before saturation, $a_{\tiny \mbox{stop}}$ is independent of $R$ for $R$ varying by more than a factor of three and with corresponding masses varying by more than an order of magnitude.  Although for clarity, the the full data set is not shown in Fig.~\ref{astop_variation}(b), the independence of $a_{\tiny \mbox{stop}}$ on $R$ and $m$ holds for our entire steel sphere data set in which $m$ varies by more than two orders of magnitude. Plotting $aR/g$ vs.\ $d$ collapses the data onto a single curve indicating that $a_{\tiny \mbox{stop}}$ varies inversely as the sphere radius.  

\begin{figure}[h!tb]
\begin{center}
\includegraphics[width=2.5in]{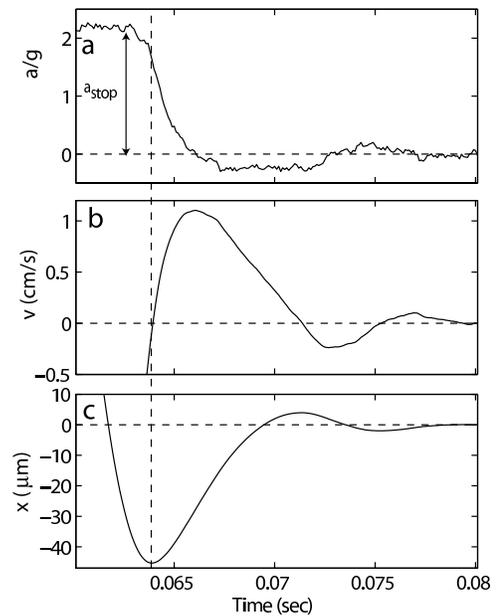}
\caption{The end of the collision (after $a_{\mathrm{stop}}$) displays (a) overshoot in acceleration which leads to (b) reversal of the collision velocity and (c) rebound in position for an $R=1.91$~cm bronze sphere colliding with glass beads at $v_c=2.4$~m/sec. (Temporal axis is zero at start of collision.)}
\label{velocity_positive}
\end{center}
\end{figure}

\begin{figure}[h!tb]
\begin{center}
\includegraphics[width=3.2in]{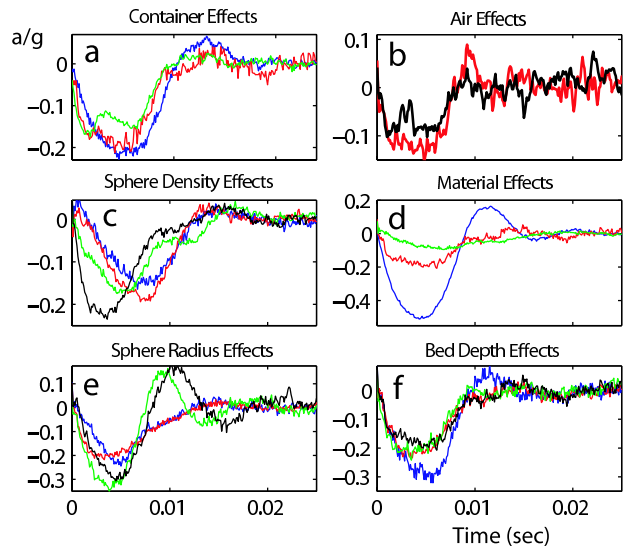}
\caption{The dependence of the final stage of the collision (see Fig.~\ref{velocity_positive}) on system parameters.   (a) Collisions averaged over interval with $2 < v_c < 5$~m/s in a cardboard barrel (blue), a PVC bucket (red), or an aluminum pot (green) (see text for container details). (b) $R=2.0$~cm sphere with $v_c=1.35$~m/s dropped into a 7~cm deep bed at atmospheric pressure (black) and evacuated to less than 50~mTorr (red).  Data is an average of 6 collisions for each pressure.  (c) $R=1.9$~cm nylon sphere colliding with $\rho_s=1.88$, 2.85, 5.03, and 9.38~grams/cm$^3$ (blue, red, green, black)  Data is an average over all collisions with $2 < v_c < 5$~m/s.  (d) Collisions with bird seed, glass beads, and cut aluminum wire in aluminum pot (blue, red, green, black). (e) Steel spheres  $R=1.3,$ 2.0, 3.5, and 4.5~cm (blue, red, green, black). (f) Layer depths of 8, 10, 15, and 20~cm (blue, red, green, black).  Glass beads are the granular medium in  (a-c,e,f), an $R=1.9$~cm bronze sphere is the impactor in (a, d-f), and the plastic bucket holds the grains in (c, e, f). }
\label{oscillationvary}
\end{center}
\end{figure}

In the final stage of the collision in the vicinity of $a_{\mathrm{stop}}$, the sphere responds as if it were in contact with an elastic-like medium, see Fig.~\ref{velocity_positive}. The sphere does not instantaneously come to rest at $v=0$ as predicted by models, but instead the acceleration decreases rapidly (with finite slope) and overshoots (negative acceleration). When the rapid decrease in $a(t)$ associated with $a_{\tiny \mbox{stop}}$ begins,  the velocity increases through zero, indicating that the sphere has reversed direction and is moving upwards; for some impact parameters (see curves Fig.~\ref{velocity_positive}(d,e)), the velocity and position can oscillate for a few cycles after the initial overshoot. Recent work by Durian's group~\cite{katAdur07} also observed an oscillation in the velocity of a sphere at the end of the collision. They attributed this to displacement of the bottom of the container, but for our data, as seen in Fig.~\ref{oscillationvary}(a), the primary overshoot is largely independent of container size and composition for three different containers: cardboard (blue), PVC (red), and aluminum (green) (see Table \ref{containers} for container details).   Nor does the interstitial air diminish the effect as Fig.~\ref{oscillationvary}(b) shows for glass beads.  Other evidence pointing to an intrinsic origin is provided by Fig.~\ref{oscillationvary}(c) which shows that as the sphere density increases the duration of the overshoot {\em decreases} and by Fig.~\ref{oscillationvary}(d) which shows that with the same impactor and container the  overshoot varies with the granular material. Additionally, the overshoot is observed to decrease with increasing sphere radius (see Fig.~\ref{oscillationvary}(e)) but to be largely independent of bed depth for depths ranging from 8~cm to 20~cm.  We speculate that when the acceleration reaches $a_{\mathrm{stop}}$, the material suddenly undergoes a solidification transition and the subsequent dynamics are a result of the sphere oscillating within the now elastic-solid as Fig.~\ref{velocity_positive}(c) suggests. This picture does not explain why the characteristic overshoot time decreases with increasing sphere density.

\subsection{Discussion of force laws}
\label{forceLaws}

\begin{figure*}[h!tb]
\begin{center}
\includegraphics[width=5in]{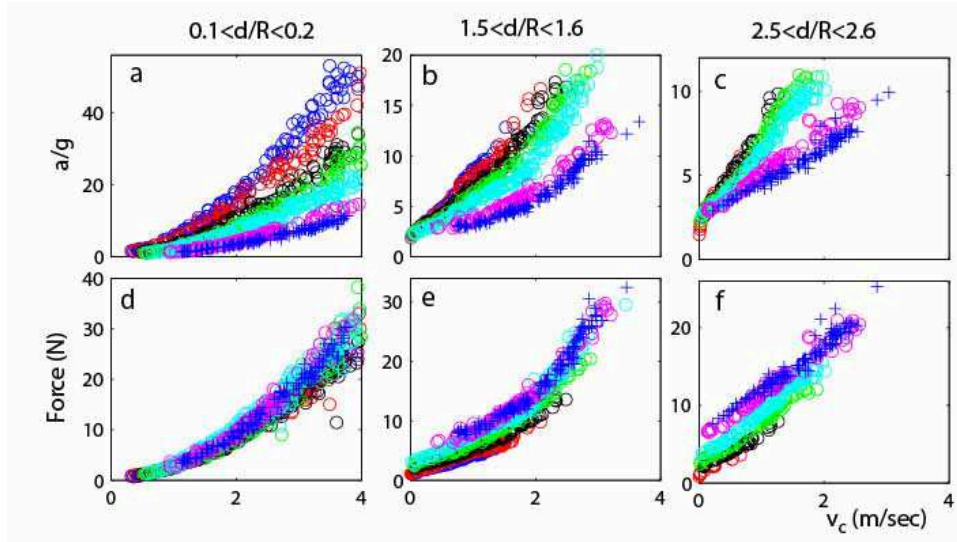}
\caption{(Color online) The acceleration (top panels (a)-(c)) and the force $m(a-g)$ (bottom panels (d)-(f)) of an $R=1.91$~cm nylon sphere with different effective densities vs.\ velocity at three distinct scaled depths $d/R$ during impact. Depths are written above each column and correspond to (a,d) initial impact, (b,e) penetration of $\approx 1.5$ sphere radii, and (c,f) penetration of $\approx 2.5$ sphere radii. Sphere densities $\rho_s=1.88,$ 2.08, 2.85, 3.91, 5.03, 7.97, and 9.38~grams/cm$^3$ correspond to blue, red, black, green, cyan, magenta $\circ$ and blue $+$.}
\label{forcelawchangedensity}
\end{center}
\end{figure*}

\begin{figure}[h!tb]
\begin{center}
\includegraphics[width=3in]{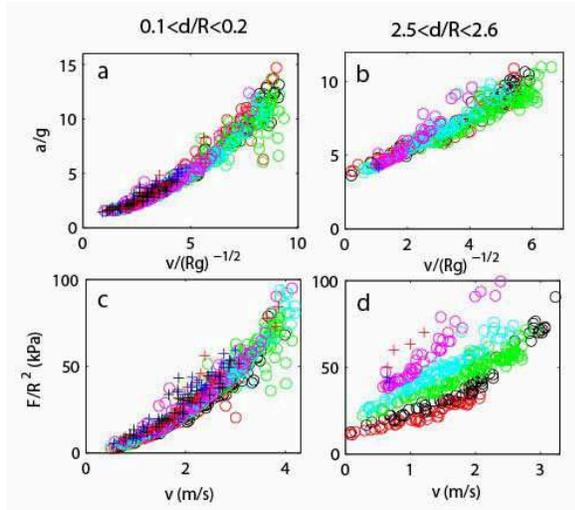}
\caption{(Color online) The acceleration (top panels (a),(b)) and the force $m(a-g)$ (bottom panels (c),(d)) of steel spheres of different radii at three scaled depths $d/R$ during impact. Depths are written above each column and correspond to (a,b) initial impact, and (c,d) penetration of $\approx 2.5$ radii. Sphere radii are $R=0.95,$ 1.27, 1.51, 1.98, 2.46, 3.49, 3.97, 4.52 and 5.00~cm with corresponding masses, $m=34.23,$ 66.3, 112, 287, 531, 1437, 2099, 3055 and 4079~grams (blue, red, black, green, cyan, magenta $\circ$ and blue, red, black $+$).}
\label{forcelawchangeradii}
\end{center}
\end{figure}

\begin{figure}[h!tb]
\begin{center}
\includegraphics[width=2.5in]{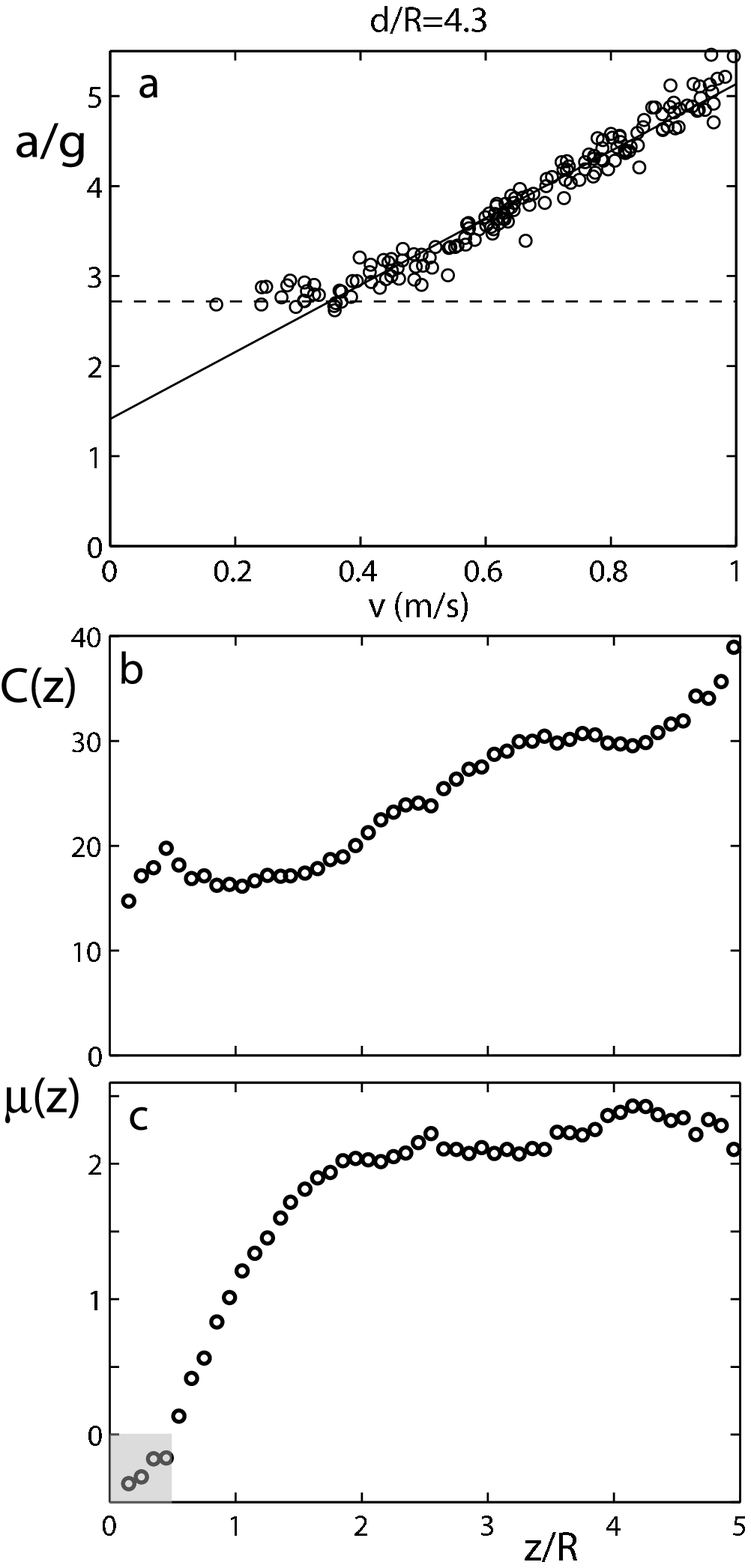}
\caption{(a) The acceleration of an $R=1.9$~cm bronze sphere vs.\ velocity at final stages of collision with glass beads (fixed depth $d/R=4.3$). The dashed line shows how $a_{\tiny \mbox{stop}}$ is determined and the solid line shows the linear fit region for this fixed depth. Compare to Fig.~\ref{forcelawchangedensity}(c) and Fig.~\ref{forcelawchangedensity}(b). (b,c) The coefficients in Eq. ~\ref{lowvelforce} vs.\ penetration depth for impact of an $R=1.9$~cm bronze sphere into glass beads are obtained from such fits. The fit region (see Fig.~\ref{forcelawchangedensity}) is for $0.25<v<1.25$~m/sec. The $\mu < 0$ region (shaded) in (c) is a result of applying a linear fit when the $v^2$ term is dominant.}
\label{coefficients}
\end{center}
\end{figure}

The preceding results for $a_{\mathrm{peak}}$ and $a_{\tiny \mbox{stop}}$ suggest that hydrodynamic forces scaling like $v^2$ dominate in the high velocity/shallow penetration regime, while frictional/hydrostatic forces are of primary importance as $v \rightarrow 0$.  To examine these ideas, Figs.~\ref{forcelawchangedensity} and \ref{forcelawchangeradii} show the velocity dependence of sphere dynamics at various fixed depths during the collision. For shallow depths and varying densities, the accelerations are different but the forces exerted by the grains, $F=m(a-g)$, collapse onto a master curve such that $F \sim v^2$ (see Fig.~\ref{forcelawchangedensity}a,d).  Thus, as assumed by ~\cite{allAmay57a,allAmay57b,tsiAvol05,ambAkam05,forAluk92,bogAdra96b}, a drag force proportional to the velocity squared, and independent of mass, is a good approximation for shallow depths.  Additionally, \cite{allAmay57a,allAmay57b,tsiAvol05,ambAkam05,forAluk92,bogAdra96b} assume the force is inversely related to the sphere cross-sectional area.  To check this scaling we plot in Fig.~\ref{forcelawchangeradii}(a,c)  $a/g$ vs.\ $v/\sqrt{Rg}$ and  $F/R^2$ vs.\ $v$ at shallow depth and for varying radii.  Both quantities fall onto master curves varying like $v^2$, indicating that indeed $F \propto R^2.$

For deeper penetration and decreasing velocity, the force and acceleration no longer vary as $v^2$, but are instead linear in velocity with a non-zero offset at $v=0$, see Figs.~\ref{forcelawchangedensity}(c,f) and ~\ref{forcelawchangeradii}(b,d).   The slope of $F$ vs.\ $v$ is independent of mass, see Fig.\ref{forcelawchangedensity}(f), suggesting a low Reynolds number fluid-like drag.  However, as indicated by Fig.~\ref{forcelawchangeradii}(b,d) the linear velocity coefficient of the force varies as $R^{5/2}$ rather than $R$ as is the case for Newtonian fluids. Linear velocity dependence has been proposed by ~\cite{allAmay57a,allAmay57b,debAwal04}, although $v_c$ was much larger ($v_c \sim 700$~m/sec in~\cite{allAmay57a,allAmay57b}) than in our study. For both varying density and radii, the extrapolated $v=0$ intercept occurs at constant acceleration indicating a force dominated by friction.  We note that at very low velocities ($v \lesssim 0.25$~m/s) $a$ becomes constant - this is the regime of $a_{\tiny \mbox{stop}}$ (see Fig.\ \ref{coefficients}(a) for example) and is why $a_{\tiny \mbox{stop}} \geqq \mu$.

Summarizing the results of Figs.~\ref{forcelawchangedensity} and ~\ref{forcelawchangeradii}, we write the drag force $F_d$ exerted by the granular medium on the sphere as the empirical force law,
\begin{equation}
F_d = \mu(z) m g + C(z) R^{5/2} \rho_s \sqrt{g} v + \alpha' R^2 v^2,
\label{lowvelforce}
\end{equation}
where $\mu(z)$ is a depth dependent constant analogous to a friction, $C(z)$ is a drag coefficient also dependent on depth, and $\alpha'$ is a constant independent of depth. We consider this equation to be valid in the ``steady" collision regime before the sidewalls of the crater start to collapse onto the impacting sphere ({\em i.e.} before the sudden jump in $a$ at $a_{\tiny \mbox{stop}}$).  The variation of $\mu$ and $C$ with depth is shown in Fig.~\ref{coefficients}. The saturation of $\mu(z)$ with penetration depth in the velocity independent term in Eq.~\ref{lowvelforce} is in accord with models of~\cite{tsiAvol05,ambAkam05}.  However, as we mention in our discussion of $a_{\tiny \mbox{stop}}$, this saturation appears to be associated with a constant effective contact area between the sphere and grains rather than a hydrostatic pressure since $\mu$ is independent of mass [{\em e.g.} see Figs. \ref{forcelawchangedensity}(c) and \ref{forcelawchangeradii}(b)].  In Eq.~\ref{lowvelforce} the linear velocity term follows the model in~\cite{debAwal04} for the scaling of $F_d$ with $R$ and $g$.  However, we find that the data scale better as $\rho_s$ while ~\cite{debAwal04} report that the scaling goes as $\sqrt {\rho_s}$.

\begin{figure}[h!tb]
\begin{center}
\includegraphics[width=3in]{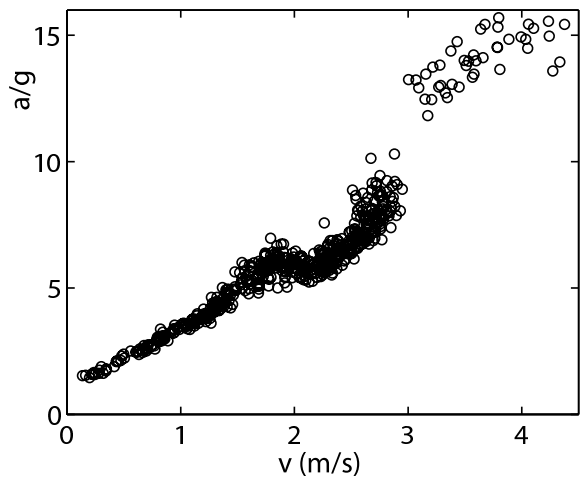}
\caption{The acceleration of an $R=1.9$~cm bronze sphere vs.\ velocity at intermediate stages of collision with glass beads (fixed depth $1.85 < d/R < 1.91$) shows non-monotonic behavior (compare to Figs.~\ref{peak_accel} and \ref{astop_variation}).}
\label{intermediateaccel}
\end{center}
\end{figure}

Finally, despite its agreement with our data in the high velocity/shallow depth and low velocity/deep depth, the empirical model proposed in Eq.~\ref{lowvelforce} is incomplete: the force in the intermediate stages of collision exhibits a more complicated non-monotonic dependence on $v(t)$ as seen in Fig.~\ref{intermediateaccel}.

\subsection{Fluctuations and Material Dependence}

\begin{figure*}[h!tb]
\begin{center}
\includegraphics[width=6in]{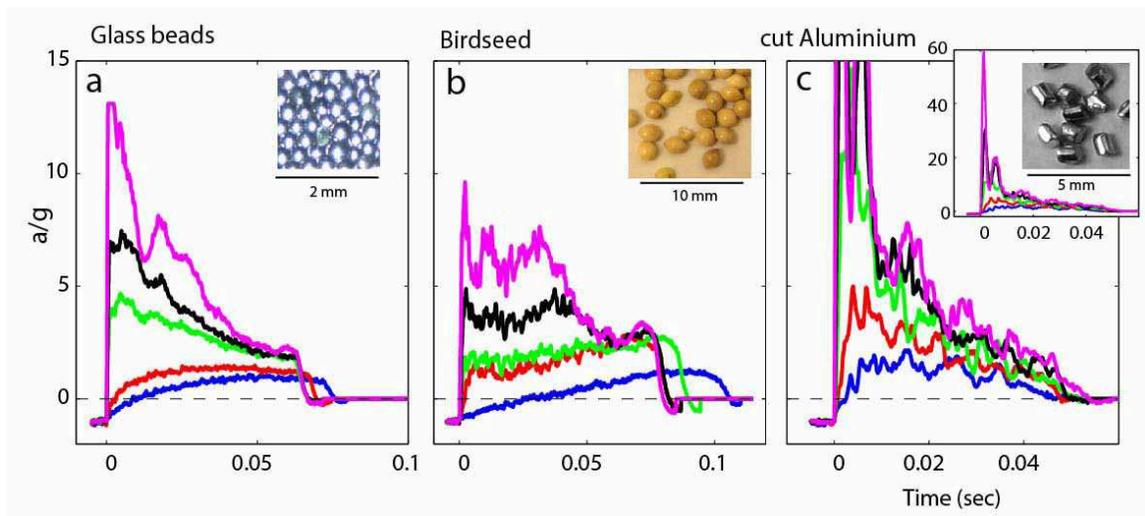}
\caption{The fluctuations in acceleration during penetration depend on the characteristics of the material being penetrated. An $R=1.9$~cm, $m=200$ gram bronze sphere impacting (a) glass spheres, (b) birdseed, and (c) cut Aluminium wire (see Table 1 for material properties).  Impact velocities (blue, red, green,black magenta) in (a) are $v_c=0.45,0.84,1.82,2.47,3.40$~m/sec, (b)$v_c=0.5,1.22,1.74,2.51,3.53$~m/sec, (c) $v_c=0.50,0.98,1.79,2.43,3.36$~m/sec.}
\label{flucsmaterial}
\end{center}
\end{figure*}

For sufficiently low impact velocity, the model of~\cite{ambAkam05} captures the shape of the acceleration profile, see Figs.~\ref{AccProfileFitComp} and \ref{modelfitdeviation}. As $v_c$ increases, the relative difference between the experimental data and the model prediction increases. For low velocity, the acceleration is concave down. According to the model, the force on the sphere increases as the sphere penetrates into the medium, and then smoothly levels out to a constant. As $v_c$ increases, the curves develop upward concavity. Substantial fluctuations in our experiments appear in the vicinity of the change in curvature which implies that they are associated with penetration dynamics dominated by the inertial $v^2$ term. The models discussed in this paper cannot capture such physics as they are purely hydrodynamic. In addition, the fluctuations depend on the type of material that the sphere impacts, see Fig.~\ref{flucsmaterial}.   The fluctuations are significantly more irregular and occur over shorter time scales in the millet seeds (smallest density) and the aluminum (largest density) than in the glass beads which exhibit a characteristic structure for higher $v_c$ (see also Fig.~\ref{intermediateaccel}).  We attribute the fluctuations in acceleration to creation and annihilation of elements of the force network~\cite{howAbeh} and are apparently strongly influenced by particle shape and also size relative to the impactor. Such fluctuations have been observed in many systems but typically in a quasi-static regime ~\cite{milAohe}.

As Fig.\ref{flucsmaterial} also shows, $a_{\mathrm{peak}}$ increases with the particle density $\rho_g$ such that the densest material (aluminum) has a peak acceleration approximately six times that of the millet seeds (birdseed) at a given $v_c$; the density ratio of the particles is approximately a factor of two (see Table~\ref{materials}).  However, the magnitude of $a_{\tiny \mbox{stop}}$ and the overshoot at the end of the collision are largest for the millet seed and smallest for aluminum which is opposite the behavior of $a_{\mathrm{peak}}$.  The collision time $t_c$ decreases monotonically with particle density.

\section{Conclusions}

We have directly measured the forces experienced by spheres and disks impacting granular media at collision velocities $v_c < 5$ m/sec. By integrating the measured acceleration we have deduced scaling relations for penetration depth and collision time as a function of $v_c$ and collision impactor parameters and compared them to scalings proposed in the literature. We have identified robust features of the collision dynamics ($a_{\mathrm{peak}}$ and $a_{\mathrm{stop}}$ and described how they scale with $v_c$ and intruder parameters. These features are in accord with proposed models of drag which propose both inertial and frictional drag terms. We have shown how the combination of such terms determines the surprising finding that time of collision is independent of velocity. Developing the force law for penetration empirically from our data, we have proposed a new model of the drag force on a sphere during impact which includes constant, linear and quadratic terms and have shown how the coefficients in this drag relation scale with $v_c$ and intruder parameters. We have discovered that during impact significant fluctuations in acceleration occur which are not described by any existing models of impact. Whether these material dependent fluctuations fall outside the scope of continuum dynamics remains to be determined.

Our experiments provide another example of the rich dynamics and fundamental physics found in the interaction of objects with granular media in the mixed fluid/solid regime. While there has been much progress in the theory of free granular flow (hydrodynamic regime)~\cite{jenAric, campbell,rerAbiz,bouAmoo}, models describing the mixed regime examined here are not rooted in fundamental physics but are largely phenomenological. Our data can constrain and inform development of models in this regime. We hypothesize that to accurately capture the dynamics we have observed (including fluctuations and stopping acceleration), the physics of rapid fluidization and solidification must be included. Beyond discrete simulations~\cite{picAlar}, multi-phase modelling like that proposed in ~\cite{volAtsi03} might be applicable or perhaps such features will have to be captured by statistical models.

Finally, in addition to probing challenging and unsolved problems in the physics of granular media, models describing the impact regime we have studied here are important to many areas of science and engineering. One area with increasing relevance is control of locomotion in organisms and robots~\cite{sarAkod,holAful06}.  Movement often occurs on complex media~\cite{autAbue} and there is a need to understand limb interaction with complex substrates~\cite{spaAgol}.  While the scalings of such properties as penetration depth and collision time might be largely insensitive to intruder and grain geometry, the details of the force developed during penetration certainly are, and thus theory is needed to model such effects.

\begin{acknowledgments}
We thank Roman Grigoriev and Robert J.\ Full and for their comments and discussions.  We also thank David Sweeney, Homero Lara, and Mateo Garcia for collecting data as well as Lionel London for photographic assistance.

\end{acknowledgments}


\begin{thebibliography}{32}
\expandafter\ifx\csname natexlab\endcsname\relax\def\natexlab#1{#1}\fi
\expandafter\ifx\csname bibnamefont\endcsname\relax
  \def\bibnamefont#1{#1}\fi
\expandafter\ifx\csname bibfnamefont\endcsname\relax
  \def\bibfnamefont#1{#1}\fi
\expandafter\ifx\csname citenamefont\endcsname\relax
  \def\citenamefont#1{#1}\fi
\expandafter\ifx\csname url\endcsname\relax
  \def\url#1{\texttt{#1}}\fi
\expandafter\ifx\csname urlprefix\endcsname\relax\def\urlprefix{URL }\fi
\providecommand{\bibinfo}[2]{#2}
\providecommand{\eprint}[2][]{\url{#2}}

\bibitem[{\citenamefont{Melosh}(1989)}]{meloshbook}
\bibinfo{author}{\bibfnamefont{H.~J.} \bibnamefont{Melosh}},
  \emph{\bibinfo{title}{Impact Cratering: A Geologic Process}}
  (\bibinfo{publisher}{Oxford University Press}, \bibinfo{year}{1989}).

\bibitem[{\citenamefont{Burrows and Hoyle}(1973)}]{burAhoy73}
\bibinfo{author}{\bibfnamefont{M.}~\bibnamefont{Burrows}} \bibnamefont{and}
  \bibinfo{author}{\bibfnamefont{G.}~\bibnamefont{Hoyle}},
  \bibinfo{journal}{Journal of Experimental Biology}
  \textbf{\bibinfo{volume}{58}}, \bibinfo{pages}{327} (\bibinfo{year}{1973}).

\bibitem[{\citenamefont{Robins}(1742)}]{robins}
\bibinfo{author}{\bibfnamefont{B.}~\bibnamefont{Robins}},
  \emph{\bibinfo{title}{New Principles of Gunnery}} (\bibinfo{year}{1742}).

\bibitem[{\citenamefont{Ciamarra et~al.}(2004)\citenamefont{Ciamarra, Lara,
  Lee, Goldman, Vishik, and Swinney}}]{picAlar}
\bibinfo{author}{\bibfnamefont{M.~P.} \bibnamefont{Ciamarra}},
  \bibinfo{author}{\bibfnamefont{A.~H.} \bibnamefont{Lara}},
  \bibinfo{author}{\bibfnamefont{A.~T.} \bibnamefont{Lee}},
  \bibinfo{author}{\bibfnamefont{D.~I.} \bibnamefont{Goldman}},
  \bibinfo{author}{\bibfnamefont{I.}~\bibnamefont{Vishik}}, \bibnamefont{and}
  \bibinfo{author}{\bibfnamefont{H.~L.} \bibnamefont{Swinney}},
  \bibinfo{journal}{Phys. Rev. Lett} \textbf{\bibinfo{volume}{92}},
  \bibinfo{pages}{194301} (\bibinfo{year}{2004}).

\bibitem[{\citenamefont{de~Bruyn and Walsh}(2004)}]{debAwal04}
\bibinfo{author}{\bibfnamefont{J.~R.} \bibnamefont{de~Bruyn}} \bibnamefont{and}
  \bibinfo{author}{\bibfnamefont{A.~M.} \bibnamefont{Walsh}},
  \bibinfo{journal}{Can. J. Phys.} \textbf{\bibinfo{volume}{82}},
  \bibinfo{pages}{439} (\bibinfo{year}{2004}).

\bibitem[{\citenamefont{Allen et~al.}(1957{\natexlab{a}})\citenamefont{Allen,
  Mayfield, and Morrison}}]{allAmay57a}
\bibinfo{author}{\bibfnamefont{W.~A.} \bibnamefont{Allen}},
  \bibinfo{author}{\bibfnamefont{E.~B.} \bibnamefont{Mayfield}},
  \bibnamefont{and} \bibinfo{author}{\bibfnamefont{H.~L.}
  \bibnamefont{Morrison}}, \bibinfo{journal}{Journal of Applied Physics}
  \textbf{\bibinfo{volume}{28}}, \bibinfo{pages}{370}
  (\bibinfo{year}{1957}{\natexlab{a}}).

\bibitem[{\citenamefont{Forrestal and Luk}(1992)}]{forAluk92}
\bibinfo{author}{\bibfnamefont{M.~J.} \bibnamefont{Forrestal}}
  \bibnamefont{and} \bibinfo{author}{\bibfnamefont{V.~K.} \bibnamefont{Luk}},
  \bibinfo{journal}{International Journal of Impact Engineering}
  \textbf{\bibinfo{volume}{12}}, \bibinfo{pages}{427} (\bibinfo{year}{1992}).

\bibitem[{\citenamefont{Lohse et~al.}(2004{\natexlab{a}})\citenamefont{Lohse,
  Bergmann, Mikkelsen, Zeilstra, van~der Meer, Versluis, van~der Weele, van~der
  Hoef, and Kuipers}}]{lohAber04}
\bibinfo{author}{\bibfnamefont{D.}~\bibnamefont{Lohse}},
  \bibinfo{author}{\bibfnamefont{R.}~\bibnamefont{Bergmann}},
  \bibinfo{author}{\bibfnamefont{R.}~\bibnamefont{Mikkelsen}},
  \bibinfo{author}{\bibfnamefont{C.}~\bibnamefont{Zeilstra}},
  \bibinfo{author}{\bibfnamefont{D.}~\bibnamefont{van~der Meer}},
  \bibinfo{author}{\bibfnamefont{M.}~\bibnamefont{Versluis}},
  \bibinfo{author}{\bibfnamefont{K.}~\bibnamefont{van~der Weele}},
  \bibinfo{author}{\bibfnamefont{M.}~\bibnamefont{van~der Hoef}},
  \bibnamefont{and} \bibinfo{author}{\bibfnamefont{H.}~\bibnamefont{Kuipers}},
  \bibinfo{journal}{Phys. Rev. Lett.} \textbf{\bibinfo{volume}{93}},
  \bibinfo{pages}{198003} (\bibinfo{year}{2004}{\natexlab{a}}).

\bibitem[{\citenamefont{Hou et~al.}(2005)\citenamefont{Hou, Peng, Liu, Lu, and
  Chan}}]{houApen05}
\bibinfo{author}{\bibfnamefont{M.}~\bibnamefont{Hou}},
  \bibinfo{author}{\bibfnamefont{Z.}~\bibnamefont{Peng}},
  \bibinfo{author}{\bibfnamefont{R.}~\bibnamefont{Liu}},
  \bibinfo{author}{\bibfnamefont{K.}~\bibnamefont{Lu}}, \bibnamefont{and}
  \bibinfo{author}{\bibfnamefont{C.~K.} \bibnamefont{Chan}},
  \bibinfo{journal}{Physical Review E} \textbf{\bibinfo{volume}{72}}
  (\bibinfo{year}{2005}), \bibinfo{note}{part 1}.

\bibitem[{\citenamefont{Boguslavskii et~al.}(1996)\citenamefont{Boguslavskii,
  Drabkin, and Salman}}]{bogAdra96b}
\bibinfo{author}{\bibfnamefont{Y.}~\bibnamefont{Boguslavskii}},
  \bibinfo{author}{\bibfnamefont{S.}~\bibnamefont{Drabkin}}, \bibnamefont{and}
  \bibinfo{author}{\bibfnamefont{A.}~\bibnamefont{Salman}},
  \bibinfo{journal}{Journal of Physics D-Applied Physics}
  \textbf{\bibinfo{volume}{29}}, \bibinfo{pages}{905} (\bibinfo{year}{1996}).

\bibitem[{\citenamefont{Wada et~al.}(2006)\citenamefont{Wada, Senshu, and
  Matsui}}]{wadAsen06}
\bibinfo{author}{\bibfnamefont{K.}~\bibnamefont{Wada}},
  \bibinfo{author}{\bibfnamefont{H.}~\bibnamefont{Senshu}}, \bibnamefont{and}
  \bibinfo{author}{\bibfnamefont{T.}~\bibnamefont{Matsui}},
  \bibinfo{journal}{Icarus} \textbf{\bibinfo{volume}{180}},
  \bibinfo{pages}{528} (\bibinfo{year}{2006}).

\bibitem[{\citenamefont{Allen et~al.}(1957{\natexlab{b}})\citenamefont{Allen,
  Mayfield, and Morrison}}]{allAmay57b}
\bibinfo{author}{\bibfnamefont{W.~A.} \bibnamefont{Allen}},
  \bibinfo{author}{\bibfnamefont{E.~B.} \bibnamefont{Mayfield}},
  \bibnamefont{and} \bibinfo{author}{\bibfnamefont{H.~L.}
  \bibnamefont{Morrison}}, \bibinfo{journal}{Journal of Applied Physics}
  \textbf{\bibinfo{volume}{28}}, \bibinfo{pages}{1331}
  (\bibinfo{year}{1957}{\natexlab{b}}).

\bibitem[{\citenamefont{Tsimring and Volfson}(2005)}]{tsiAvol05}
\bibinfo{author}{\bibfnamefont{L.~S.} \bibnamefont{Tsimring}} \bibnamefont{and}
  \bibinfo{author}{\bibfnamefont{D.}~\bibnamefont{Volfson}}, in
  \emph{\bibinfo{booktitle}{Powders and Grains 2005}}, edited by
  \bibinfo{editor}{\bibfnamefont{R.}~\bibnamefont{Garc\'{i}a-Rojo}},
  \bibinfo{editor}{\bibfnamefont{H.~J.} \bibnamefont{Herrmann}},
  \bibnamefont{and} \bibinfo{editor}{\bibfnamefont{S.}~\bibnamefont{McNamara}}
  (\bibinfo{year}{2005}), vol.~\bibinfo{volume}{2}, pp.
  \bibinfo{pages}{1215--1223}.

\bibitem[{\citenamefont{Ambroso
  et~al.}(2005{\natexlab{a}})\citenamefont{Ambroso, Kamien, and
  Durian}}]{ambAkam05}
\bibinfo{author}{\bibfnamefont{M.~A.} \bibnamefont{Ambroso}},
  \bibinfo{author}{\bibfnamefont{R.~D.} \bibnamefont{Kamien}},
  \bibnamefont{and} \bibinfo{author}{\bibfnamefont{D.~J.}
  \bibnamefont{Durian}}, \bibinfo{journal}{Phys. Rev. E}
  \textbf{\bibinfo{volume}{72}}, \bibinfo{pages}{041305}
  (\bibinfo{year}{2005}{\natexlab{a}}).

\bibitem[{\citenamefont{Ambroso
  et~al.}(2005{\natexlab{b}})\citenamefont{Ambroso, Santore, Abate, and
  Durian}}]{ambAsan05}
\bibinfo{author}{\bibfnamefont{M.~A.} \bibnamefont{Ambroso}},
  \bibinfo{author}{\bibfnamefont{C.~R.} \bibnamefont{Santore}},
  \bibinfo{author}{\bibfnamefont{A.~R.} \bibnamefont{Abate}}, \bibnamefont{and}
  \bibinfo{author}{\bibfnamefont{D.~J.} \bibnamefont{Durian}},
  \bibinfo{journal}{Physical Review E} \textbf{\bibinfo{volume}{71}}
  (\bibinfo{year}{2005}{\natexlab{b}}), \bibinfo{note}{part 1}.

\bibitem[{\citenamefont{Stone et~al.}(2004)\citenamefont{Stone, Barry,
  Bernstein, Pelc, Tsui, and Schiffer}}]{stoAbar04}
\bibinfo{author}{\bibfnamefont{M.~B.} \bibnamefont{Stone}},
  \bibinfo{author}{\bibfnamefont{R.}~\bibnamefont{Barry}},
  \bibinfo{author}{\bibfnamefont{D.~P.} \bibnamefont{Bernstein}},
  \bibinfo{author}{\bibfnamefont{M.~D.} \bibnamefont{Pelc}},
  \bibinfo{author}{\bibfnamefont{Y.~K.} \bibnamefont{Tsui}}, \bibnamefont{and}
  \bibinfo{author}{\bibfnamefont{P.}~\bibnamefont{Schiffer}},
  \bibinfo{journal}{Physical Review E} \textbf{\bibinfo{volume}{70}}
  (\bibinfo{year}{2004}), \bibinfo{note}{part 1}.

\bibitem[{\citenamefont{Katsuragi and Durian}(2007)}]{katAdur07}
\bibinfo{author}{\bibfnamefont{H.}~\bibnamefont{Katsuragi}} \bibnamefont{and}
  \bibinfo{author}{\bibfnamefont{D.~J.} \bibnamefont{Durian}},
  \bibinfo{journal}{Nature Physics} \textbf{\bibinfo{volume}{3}},
  \bibinfo{pages}{420} (\bibinfo{year}{2007}).

\bibitem[{\citenamefont{Lohse et~al.}(2004{\natexlab{b}})\citenamefont{Lohse,
  Rauh\'{e}, Bergmann, and van~der Meer}}]{lohArau}
\bibinfo{author}{\bibfnamefont{D.}~\bibnamefont{Lohse}},
  \bibinfo{author}{\bibfnamefont{R.}~\bibnamefont{Rauh\'{e}}},
  \bibinfo{author}{\bibfnamefont{R.}~\bibnamefont{Bergmann}}, \bibnamefont{and}
  \bibinfo{author}{\bibfnamefont{D.}~\bibnamefont{van~der Meer}},
  \bibinfo{journal}{Nature} \textbf{\bibinfo{volume}{432}},
  \bibinfo{pages}{689} (\bibinfo{year}{2004}{\natexlab{b}}).

\bibitem[{\citenamefont{Thoroddsen and Shen}(2001)}]{thoAshe}
\bibinfo{author}{\bibfnamefont{S.~T.} \bibnamefont{Thoroddsen}}
  \bibnamefont{and} \bibinfo{author}{\bibfnamefont{A.~Q.} \bibnamefont{Shen}},
  \bibinfo{journal}{Phys. Fluids} \textbf{\bibinfo{volume}{13}},
  \bibinfo{pages}{4} (\bibinfo{year}{2001}).

\bibitem[{\citenamefont{Nedderman}(1992)}]{nedderman}
\bibinfo{author}{\bibfnamefont{R.~M.} \bibnamefont{Nedderman}},
  \emph{\bibinfo{title}{Statics and kinematics of granular materials}}
  (\bibinfo{publisher}{Cambridge University Press},
  \bibinfo{address}{Cambridge}, \bibinfo{year}{1992}).

\bibitem[{\citenamefont{Tritton}(1989)}]{tritton89}
\bibinfo{author}{\bibfnamefont{D.}~\bibnamefont{Tritton}},
  \emph{\bibinfo{title}{Physical Fluid Dynamics}} (\bibinfo{publisher}{Oxford
  University Press}, \bibinfo{year}{1989}).

\bibitem[{\citenamefont{Howell et~al.}(1999)\citenamefont{Howell, Behringer,
  and Veje}}]{howAbeh}
\bibinfo{author}{\bibfnamefont{D.}~\bibnamefont{Howell}},
  \bibinfo{author}{\bibfnamefont{R.~P.} \bibnamefont{Behringer}},
  \bibnamefont{and} \bibinfo{author}{\bibfnamefont{C.}~\bibnamefont{Veje}},
  \bibinfo{journal}{Phys. Rev. Lett} \textbf{\bibinfo{volume}{82}},
  \bibinfo{pages}{5241} (\bibinfo{year}{1999}).

\bibitem[{\citenamefont{Miller et~al.}(1996)\citenamefont{Miller, O'Hern, and
  Behringer}}]{milAohe}
\bibinfo{author}{\bibfnamefont{B.}~\bibnamefont{Miller}},
  \bibinfo{author}{\bibfnamefont{C.}~\bibnamefont{O'Hern}}, \bibnamefont{and}
  \bibinfo{author}{\bibfnamefont{R.~P.} \bibnamefont{Behringer}},
  \bibinfo{journal}{Phys. Rev. Lett} \textbf{\bibinfo{volume}{77}},
  \bibinfo{pages}{3110} (\bibinfo{year}{1996}).

\bibitem[{\citenamefont{Jenkins and Richman}(1985)}]{jenAric}
\bibinfo{author}{\bibfnamefont{J.}~\bibnamefont{Jenkins}} \bibnamefont{and}
  \bibinfo{author}{\bibfnamefont{M.}~\bibnamefont{Richman}},
  \bibinfo{journal}{Arch. Rat. Mech. Anal.} \textbf{\bibinfo{volume}{87}},
  \bibinfo{pages}{355} (\bibinfo{year}{1985}).

\bibitem[{\citenamefont{Campbell}(1990)}]{campbell}
\bibinfo{author}{\bibfnamefont{C.~S.} \bibnamefont{Campbell}},
  \bibinfo{journal}{Annu. Rev. Fluid Mech.} \textbf{\bibinfo{volume}{2}},
  \bibinfo{pages}{57} (\bibinfo{year}{1990}).

\bibitem[{\citenamefont{Rericha et~al.}(2001)\citenamefont{Rericha, Bizon,
  Shattuck, and Swinney}}]{rerAbiz}
\bibinfo{author}{\bibfnamefont{E.~C.} \bibnamefont{Rericha}},
  \bibinfo{author}{\bibfnamefont{C.}~\bibnamefont{Bizon}},
  \bibinfo{author}{\bibfnamefont{M.~D.} \bibnamefont{Shattuck}},
  \bibnamefont{and} \bibinfo{author}{\bibfnamefont{H.~L.}
  \bibnamefont{Swinney}}, \bibinfo{journal}{Phys. Rev. Lett.}
  \textbf{\bibinfo{volume}{88}}, \bibinfo{pages}{014302}
  (\bibinfo{year}{2001}).

\bibitem[{\citenamefont{Bougie et~al.}(2002)\citenamefont{Bougie, Moon, Swift,
  and Swinney}}]{bouAmoo}
\bibinfo{author}{\bibfnamefont{J.}~\bibnamefont{Bougie}},
  \bibinfo{author}{\bibfnamefont{S.~J.} \bibnamefont{Moon}},
  \bibinfo{author}{\bibfnamefont{J.~B.} \bibnamefont{Swift}}, \bibnamefont{and}
  \bibinfo{author}{\bibfnamefont{H.~L.} \bibnamefont{Swinney}},
  \bibinfo{journal}{Phys. Rev. E} \textbf{\bibinfo{volume}{66}},
  \bibinfo{pages}{051301} (\bibinfo{year}{2002}).

\bibitem[{\citenamefont{Volfson et~al.}(2003)\citenamefont{Volfson, Tsimring,
  and Aranson}}]{volAtsi03}
\bibinfo{author}{\bibfnamefont{D.}~\bibnamefont{Volfson}},
  \bibinfo{author}{\bibfnamefont{L.~S.} \bibnamefont{Tsimring}},
  \bibnamefont{and} \bibinfo{author}{\bibfnamefont{I.~S.}
  \bibnamefont{Aranson}}, \bibinfo{journal}{Physical Review E}
  \textbf{\bibinfo{volume}{68}}, \bibinfo{pages}{021301}
  (\bibinfo{year}{2003}).

\bibitem[{\citenamefont{Saranli and Koditschek}(2003)}]{sarAkod}
\bibinfo{author}{\bibfnamefont{U.}~\bibnamefont{Saranli}} \bibnamefont{and}
  \bibinfo{author}{\bibfnamefont{D.~E.} \bibnamefont{Koditschek}}, in
  \emph{\bibinfo{booktitle}{Proceedings of the IEEE International Conference On
  Robotics and Automation}} (\bibinfo{year}{2003}), vol.~\bibinfo{volume}{1},
  pp. \bibinfo{pages}{1374--1379}.

\bibitem[{\citenamefont{Holmes et~al.}(2006)\citenamefont{Holmes, Full,
  Koditschek, and Guckenheimer}}]{holAful06}
\bibinfo{author}{\bibfnamefont{P.}~\bibnamefont{Holmes}},
  \bibinfo{author}{\bibfnamefont{R.~J.} \bibnamefont{Full}},
  \bibinfo{author}{\bibfnamefont{D.}~\bibnamefont{Koditschek}},
  \bibnamefont{and}
  \bibinfo{author}{\bibfnamefont{J.}~\bibnamefont{Guckenheimer}},
  \bibinfo{journal}{Siam Review} \textbf{\bibinfo{volume}{48}},
  \bibinfo{pages}{207} (\bibinfo{year}{2006}).

\bibitem[{\citenamefont{Autumn et~al.}(2005)\citenamefont{Autumn, Buehler,
  Cutkosky, Fearing, Full, Goldman, Groff, Provancher, Rizzi, Saranli
  et~al.}}]{autAbue}
\bibinfo{author}{\bibfnamefont{K.}~\bibnamefont{Autumn}},
  \bibinfo{author}{\bibfnamefont{M.}~\bibnamefont{Buehler}},
  \bibinfo{author}{\bibfnamefont{M.}~\bibnamefont{Cutkosky}},
  \bibinfo{author}{\bibfnamefont{R.}~\bibnamefont{Fearing}},
  \bibinfo{author}{\bibfnamefont{R.~J.} \bibnamefont{Full}},
  \bibinfo{author}{\bibfnamefont{D.}~\bibnamefont{Goldman}},
  \bibinfo{author}{\bibfnamefont{R.}~\bibnamefont{Groff}},
  \bibinfo{author}{\bibfnamefont{W.}~\bibnamefont{Provancher}},
  \bibinfo{author}{\bibfnamefont{A.~E.} \bibnamefont{Rizzi}},
  \bibinfo{author}{\bibfnamefont{U.}~\bibnamefont{Saranli}},
  \bibnamefont{et~al.}, in \emph{\bibinfo{booktitle}{Unmanned Ground Vehicle
  Technology VII}}, edited by \bibinfo{editor}{\bibfnamefont{D.~W.~G.}
  \bibnamefont{Grant R.~Gerhart}, \bibfnamefont{Charles M.~Shoemaker}}
  (\bibinfo{year}{2005}), vol. \bibinfo{volume}{5804} of
  \emph{\bibinfo{series}{Proceedings of SPIE}}, pp. \bibinfo{pages}{291--302}.

\bibitem[{\citenamefont{Spagna et~al.}(2007)\citenamefont{Spagna, Goldman, Lin,
  Koditschek, and Full}}]{spaAgol}
\bibinfo{author}{\bibfnamefont{J.~P.} \bibnamefont{Spagna}},
  \bibinfo{author}{\bibfnamefont{D.~I.} \bibnamefont{Goldman}},
  \bibinfo{author}{\bibfnamefont{P.-C.} \bibnamefont{Lin}},
  \bibinfo{author}{\bibfnamefont{D.~E.} \bibnamefont{Koditschek}},
  \bibnamefont{and} \bibinfo{author}{\bibfnamefont{R.~J.} \bibnamefont{Full}},
  \bibinfo{journal}{Bioinspiration and Biomimetics}
  \textbf{\bibinfo{volume}{2}}, \bibinfo{pages}{9} (\bibinfo{year}{2007}).

\end{thebibliography}
\end{document}